\documentclass[aps,prl,reprint,superscriptaddress,amsmath,amssymb,floatfix]{revtex4-2}
\usepackage{graphicx}
\usepackage{bm}
\usepackage[colorlinks=true,linkcolor=blue,citecolor=blue,urlcolor=blue]{hyperref}

\newcommand{\sgn}{\operatorname{sign}}

\begin{document}

\title{Reflection Zeros Set the Scattering Invariant of Finite Majorana Wires}

\author{Alex Krasnok}
\email{akrasnok@fiu.edu}
\affiliation{Department of Electrical and Computer Engineering, Florida International University, Miami, Florida 33174, USA}

\begin{abstract}
Majorana modes, zero-energy excitations of superconducting wires that are their own antiparticles, could store quantum information protected from local noise. In transport experiments, however, ordinary bound states mimic their signature, a conductance peak at zero voltage, and whether Majorana modes have been observed remains debated. A sign, $Q$, of the electron reflection at zero energy marks the topological phase of a long wire, but what $Q$ certifies in a finite device has been unclear. Here we show for the first time that $Q$ counts reflectionless states, waves that enter through one contact and leave entirely through the other: $Q=-1$ when an odd number of them lie in the lower half of the complex energy plane. In a finite wire this means that the two Majorana modes are split by less than their coupling to the contacts. With equal contacts, every sign change of $Q$, including the bulk topological transition, is an exceptional point at which two such states merge and transmission is maximally flat. At zero temperature, $Q=-1$ exactly when the zero-bias conductance measured at a contact exceeds $e^2/h$. These results show what transport can certify about a Majorana device and how to read the invariant from conductance.
\end{abstract}

\maketitle

A Majorana mode is a zero-energy excitation of a superconductor that is its own antiparticle. Two such modes at the ends of a topological superconducting wire share a single fermionic state whose occupation no local perturbation can detect or change, and this makes them candidates for protected qubits~\cite{ReadGreen2000,Kitaev2001,Nayak2008,Alicea2012}. Proposals based on topological insulators and semiconductor nanowires~\cite{FuKane2008,Lutchyn2010,Oreg2010} were followed by the observation of the expected signature, a conductance peak at zero bias voltage~\cite{Mourik2012}. However, ordinary bound states of nontopological (trivial) wires, among them the partially separated pairs known as quasi-Majorana states, produce similar peaks~\cite{Kells2012,Liu2017,Moore2018,Vuik2019,Prada2020}. The same ambiguity runs through the debate over semiconductor-superconductor (InAs/Al) devices that pass the topological gap protocol, a sequence of conductance tests~\cite{Pikulin2021,Aghaee2023TGP,Microsoft2025Nature,Legg2025TGP,Legg2026Nature,Aghaee2025Response}, and through the minimal Kitaev chains built from coupled quantum dots~\cite{LeijnseFlensberg2012,Dvir2023,tenHaaf2024,Bordin2025}. What does transport through a finite device reveal about its topology?

\begin{figure*}[t]
\includegraphics[width=\textwidth]{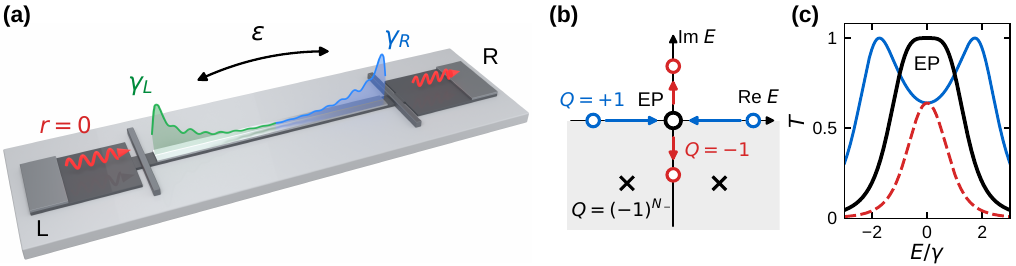}
\caption{Reflectionless states of a Majorana wire. (a) InAs/Al wire between normal leads L and R. Tunnel gates set the couplings $\gamma_L$ and $\gamma_R$ of the Majorana end modes (computed densities in green and blue), which are split by $\varepsilon$. At a real reflection zero a wave sent from L (red) leaves through R without reflection. (b) Zeros (circles) and poles (crosses, $\varepsilon=\gamma$) of $\det r_L(E)$ for $\gamma_L=\gamma_R=\gamma$. A real pair ($\varepsilon=2\gamma$, blue) coalesces at $E=0$ ($\varepsilon=\gamma$) and splits along the imaginary axis ($\varepsilon=\gamma/2$, red), which puts one zero in the lower half-plane (shaded), where $Q=-1$. (c) Transmission of the Majorana channel for the same three values of $\varepsilon$.}
\label{fig1}
\end{figure*}

For a long wire, scattering theory gives a clear answer. An electron sent from a normal lead at energy $E$, measured from the Fermi level, returns as an electron or as a hole, with amplitudes that form the reflection matrix $r(E)$. In a superconductor without spin-rotation and time-reversal symmetry (class D), the symmetry between electrons and holes (particle-hole symmetry) makes $r$ real at $E=0$, and $Q=\sgn\det r(0)$ is the topological invariant, the label that separates the two phases~\cite{Akhmerov2011,Fulga2011,Fulga2012,Beenakker2015}. It equals $-1$ when the lead couples to an unpaired Majorana mode and changes only where $\det r(0)=0$, which requires a perfectly transmitted mode~\cite{Akhmerov2011}. A finite wire is different. Probed through one lead it always has $Q=+1$~\cite{Pikulin2012,ArayaDay2025}, and between two leads its two Majorana modes give $Q=\sgn(\varepsilon^2-\gamma_L\gamma_R)$~\cite{ArayaDay2025}, with $\varepsilon$ the energy splitting caused by their overlap and $\gamma_L$ and $\gamma_R$ the rates, in energy units, at which they leak into the left and right leads. The sign that labels a phase of the long wire thus depends, in a finite one, on how the wire is contacted.

Tunneling spectroscopy sees the poles of the scattering matrix, the complex energies of decaying states (resonances), whose real and imaginary parts set the positions and widths of conductance peaks~\cite{Law2009,Flensberg2010}. Two poles can merge on the imaginary axis~\cite{Pikulin2012,Pikulin2013} at an exceptional point (EP), where two eigenvalues of a non-Hermitian operator, the effective Hamiltonian of an open system, coalesce together with their eigenvectors~\cite{Heiss2012,MiriAlu2019,Bergholtz2021,Ashida2020}, and in a trivial wire such a merger can pin a conductance peak at zero bias~\cite{SanJose2016,Avila2019}. Photonics treats the zeros of a scattering matrix, the complex energies at which chosen outgoing waves vanish, as equally physical~\cite{Grigoriev2013,Krasnok2019}. At a real zero an incident wave is absorbed completely (coherent perfect absorption)~\cite{Chong2010,BaranovNRM2017}, and the zeros of a reflection matrix define reflectionless scattering modes, states fed through some channels that send nothing back into them~\cite{Sweeney2020,Stone2021}. Two zeros can coalesce at an EP, which flattens the response~\cite{Sweeney2019,Wang2021,Ferise2022}, zero reflection can be topologically protected~\cite{Kravets2013,Soori2026}, and the reflection matrix has been treated as a non-Hermitian operator~\cite{Franca2022}. Reflectionless states were also predicted in quantum scattering~\cite{Soley2023} and in multiterminal Josephson junctions, where at zero energy they mark the boundaries of topological phases and two zeros merge there at an EP~\cite{Ohnmacht2025}. How the reflection zeros of a Majorana wire relate to its invariant $Q$ has not been established.

In this Letter, we show that $Q$ counts reflection zeros: $Q=(-1)^{N_-}$, with $N_-$ the number of zeros below the real axis, because the poles contribute a fixed sign at $E=0$. This gives the invariant of a finite device a physical meaning: every sign change of $Q$ is a reflectionless state passing through $E=0$. With mirror symmetry two zeros meet there, so each sign change is an EP of reflectionless states with maximally flat transmission, $1-T\propto E^4$. We find this EP in models of an InAs/Al wire and of a gate-defined two-dimensional device, at the bulk transition and at each later sign change. The count is also measurable. For a lead coupled through one Majorana mode the conductance measured at that lead is $G_{LL}(E)=(e^2/h)[1-\operatorname{Re}\det r_L(E)]$, so at zero temperature $Q=-1$ exactly when the zero-bias conductance exceeds $e^2/h$, and at finite temperature $Q$ follows from fits of both local spectra. To our knowledge, the counting rule and this conductance threshold are reported here for the first time. Together they tell an experiment what a measured $Q$ certifies, a Majorana splitting smaller than the coupling to the leads, and how to read it from conductance.

\textit{Zeros and poles.} Consider a wire with an electron-hole (Bogoliubov-de Gennes) Hamiltonian $H$ of even dimension $N$, attached to leads L and R through coupling matrices $W_L$ and $W_R$. For leads with an energy-independent density of states (the wide-band limit) the scattering matrix is $S(E)=1-2\pi i W^\dagger(E-H+i\Gamma/2)^{-1}W$, with $W=(W_L,W_R)$, where $\Gamma=\Gamma_L+\Gamma_R$ and $\Gamma_a=2\pi W_aW_a^\dagger$ is the matrix of escape rates into lead $a$~\cite{Mahaux1969,Beenakker2015}. We use the time dependence $e^{-iEt/\hbar}$, so resonances lie in the lower half-plane. The reflection block of lead L then obeys (Sec.~S1 of the Supplemental Material~\cite{SM})
\begin{equation}
\det r_L(E)=\frac{\det(E-H_Z)}{\det(E-H_P)},\qquad H_{P,Z}=H\mp\frac{i}{2}\Gamma_L-\frac{i}{2}\Gamma_R .
\label{eq:ratio}
\end{equation}
The poles are eigenvalues of $H_P$, in which both leads drain the wire. The zeros are eigenvalues of $H_Z$, in which lead L feeds the wire, as a gain would, and lead R drains it~\cite{Li2017,Fyodorov2017,Sweeney2020}. A reflection zero is a state that accepts everything lead L sends and releases it into lead R [Fig.~\ref{fig1}(a)]. On the real axis it is a perfectly transmitted channel, and off the axis it is reflectionless for an incident wave whose amplitude grows or decays in time. For energy-dependent leads the counting uses the coupling at $E=0$, which is exact for $Q$~\cite{SM}.

\textit{Counting rule.} The sign of $\det r_L(0)$ is fixed by the zeros alone. In a basis of self-conjugate (Majorana) operators particle-hole symmetry makes $H=iA$ with $A$ real and antisymmetric, and makes $\Gamma_L$ and $\Gamma_R$ real~\cite{Beenakker2015}. Then $H_{P,Z}=iM_{P,Z}$ with real matrices $M_P=A-(\Gamma_L+\Gamma_R)/2$ and $M_Z=A+(\Gamma_L-\Gamma_R)/2$, and Eq.~(\ref{eq:ratio}) at $E=0$ gives $\det r_L(0)=\det M_Z/\det M_P$. Because both leads drain the wire, the real eigenvalues of $M_P$ are negative, and since $N$ is even they are even in number, so $\det M_P>0$ (Sec.~S2~\cite{SM}). The pole factor never contributes a sign, and $Q=\sgn\det M_Z$. The zeros $z=i\lambda$, with $\lambda$ the eigenvalues of $M_Z$, are symmetric under $z\to-z^*$. Zeros off the imaginary axis form such pairs and contribute $|z|^2>0$ to $\det M_Z$, whereas every zero on the negative imaginary axis contributes a negative factor. Hence
\begin{equation}
Q\equiv\sgn\det r_L(0)=(-1)^{N_-},
\label{eq:count}
\end{equation}
where $N_-$ is the number of reflection zeros with $\operatorname{Im}z<0$.

Equation~(\ref{eq:count}) has two consequences. First, $Q$ changes only when a zero crosses $E=0$ along the imaginary axis, whereas a pair of zeros crossing the real axis at $\pm\operatorname{Re}z\neq0$ leaves it unchanged. This is the condition $\det r(0)=0$ of Ref.~\cite{Akhmerov2011}, stated as an eigenvalue condition on $H_Z$. Second, with a single lead ($\Gamma_R=0$) every zero is the complex conjugate of a pole, all zeros lie in the upper half-plane, and $Q=+1$ for any finite wire~\cite{Pikulin2012,ArayaDay2025}. A zero moves below the real axis only if something drains the wire, the far lead or loss inside the wire~\cite{ArayaDay2025,SM}. $Q$ also equals $(-1)^W$, with $W$ the winding number of $\det r_L(E)$ along the real axis. Lead R gives the same $Q$, and the rule extends to wires with time-reversal symmetry (Sec.~S2~\cite{SM}).

\begin{figure*}[t]
\includegraphics[width=\textwidth]{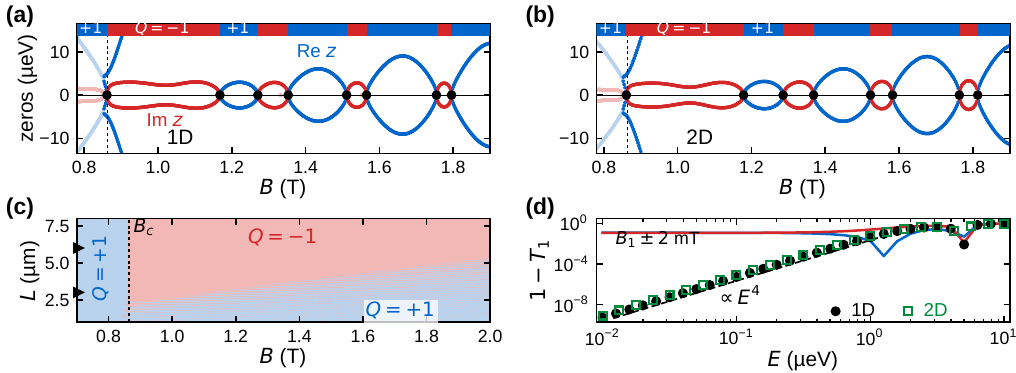}
\caption{Sign changes of $Q$ for equal couplings. Reflection zeros nearest $E=0$ versus magnetic field in (a) the InAs/Al wire (1D, $L=3~\mu$m) and (b) the gate-defined device (2D): real zeros (blue), imaginary zeros (red), and both parts of complex quartets (pale). Dots mark coalescence fields, dashed lines the bulk critical field $B_c$, and the top strips $Q$ from the scattering matrix. (c) $Q$ of the wire versus field and length; triangles mark $L=3$ and $6~\mu$m. (d) Reflection $1-T_1$ of the most transmitted channel at the first sign change $B_1$ in the wire (dots) and in the 2D device (squares), and in the wire $2$~mT below (blue) and above (red) $B_1$.}
\label{fig2}
\end{figure*}

\textit{Two Majorana modes.} When the other states of the wire are far from $E=0$, the zeros and poles nearest $E=0$ come from the two Majorana end modes~\cite{Nilsson2008}. With splitting $\varepsilon$ and escape rates (half-widths) $\gamma_L$ and $\gamma_R$ set by the leads, $A=\varepsilon\,(\begin{smallmatrix}0&1\\-1&0\end{smallmatrix})$, $\Gamma_L/2=\mathrm{diag}(\gamma_L,0)$, and $\Gamma_R/2=\mathrm{diag}(0,\gamma_R)$, so that $\det r_L(E)=(E-z_+)(E-z_-)/[(E-p_+)(E-p_-)]$ with
\begin{equation}
\begin{aligned}
z_\pm&=\tfrac{i}{2}(\gamma_L-\gamma_R)\pm\big[\varepsilon^2-\tfrac14(\gamma_L+\gamma_R)^2\big]^{1/2},\\
p_\pm&=-\tfrac{i}{2}(\gamma_L+\gamma_R)\pm\big[\varepsilon^2-\tfrac14(\gamma_L-\gamma_R)^2\big]^{1/2}.
\end{aligned}
\label{eq:twomodes}
\end{equation}
Since $\det M_Z=\varepsilon^2-\gamma_L\gamma_R$, Eq.~(\ref{eq:count}) reproduces $Q=\sgn(\varepsilon^2-\gamma_L\gamma_R)$, Eq.~(7) of Ref.~\cite{ArayaDay2025}: in a finite wire, $Q=-1$ states that the Majorana splitting is smaller than the geometric mean of the two couplings. Figure~\ref{fig1}(b) shows the zeros for $\gamma_L=\gamma_R=\gamma$. For $\varepsilon>\gamma$ they form a real pair at $\pm(\varepsilon^2-\gamma^2)^{1/2}$, two energies of perfect transmission [Fig.~\ref{fig1}(c)], and $Q=+1$. As $\varepsilon$ decreases, the pair meets at $E=0$ for $\varepsilon=\gamma$ and splits along the imaginary axis. One zero then lies in the lower half-plane, and $Q=-1$. The poles $p_\pm=\pm\varepsilon-i\gamma$ pass this point smoothly, although $Q=+1$ still means that they are farther apart than their width.

\textit{Exceptional point.} The meeting point is an EP of $H_Z$. At $\varepsilon=\gamma$, $H_Z=i\gamma\,(\begin{smallmatrix}1&1\\-1&-1\end{smallmatrix})$ squares to zero: it cannot be diagonalized and has a single eigenvector~\cite{SM}. Near this point the zeros split as $z\simeq\pm\gamma[2(\varepsilon-\gamma)/\gamma]^{1/2}$. Two symmetries pin the EP to $E=0$. Particle-hole symmetry maps a zero $z$ to $-z^*$, and a mirror operation $P$ that exchanges the two ends of the device gives $PH_ZP^{-1}=H_Z^\dagger$, which maps $z$ to $z^*$. Zeros that obey both maps form real pairs, imaginary pairs, or complex quartets, and $Q$ is the parity of the number of imaginary pairs. It can change only when a real pair turns into an imaginary one, which requires the two zeros to meet at $E=0$. In a mirror-symmetric device every sign change of $Q$ is therefore a double zero at $E=0$, whatever the number of modes~\cite{SM}, like the symmetry-protected EPs of non-Hermitian band theory~\cite{Budich2019,Okugawa2019,Kawabata2019}. On the real axis a double zero gives $|\det r_L|^2\propto E^4$, and in the two-mode model the transmission probability of the Majorana channel~\cite{SM} is
\begin{equation}
T(E)=1-|\det r_L(E)|^2=\frac{4\gamma^4}{E^4+4\gamma^4},
\label{eq:butter}
\end{equation}
the maximally flat (second-order Butterworth~\cite{Butterworth1930}) response of two critically coupled resonators, observed at reflectionless EPs of microwave cavities~\cite{Ferise2022}. The Majorana channel carries heat~\cite{Akhmerov2011,SM} but almost no charge, because an electron that enters it leaves the other lead as an electron or as a hole with equal probability~\cite{Rosdahl2018,Danon2020}. The electrical signature of the zeros is local.

\begin{figure*}[t]
\includegraphics[width=\textwidth]{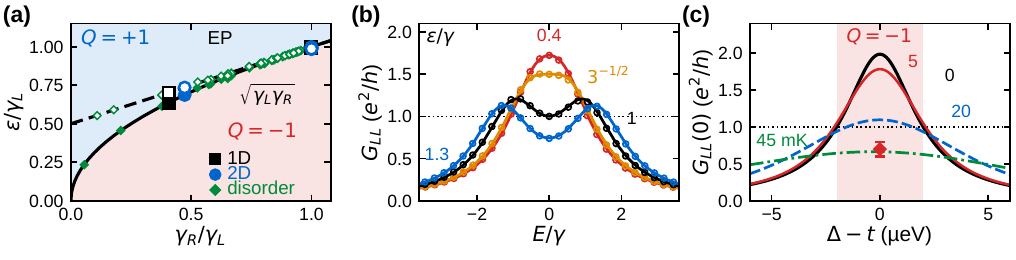}
\caption{(a) Unequal couplings separate the sign change of $Q$, at $\varepsilon=(\gamma_L\gamma_R)^{1/2}$ (solid), from the EP, at $\varepsilon=(\gamma_L+\gamma_R)/2$ (dashed). Symbols show the wire (squares), the 2D device (circles), and disordered wires (diamonds) at their sign-change fields (filled) and coalescence fields (open). (b) Local conductance of the wire at zero temperature for $\varepsilon/\gamma=0.4$, $3^{-1/2}$, $1$, and $1.3$ from the scattering matrix (symbols) and from the two-mode form of Eq.~(\ref{eq:GLL}) (lines). (c) Zero-bias conductance of a two-site Kitaev chain with the model parameters of Ref.~\cite{Dvir2023} at 0, 5, 20, and 45~mK. The diamond shows the measured peak heights~\cite{Dvir2023}. Red shading marks $Q=-1$, and dotted lines mark $e^2/h$.}
\label{fig3}
\end{figure*}

\textit{Local conductance.} When lead L couples to the wire through a single Majorana mode, $r_L$ differs from the identity by a matrix of rank one, and the local conductance $G_{LL}=dI_L/dV_L$, the response of the current in lead L to its own bias voltage, is for any number of lead channels (Sec.~S4~\cite{SM})
\begin{equation}
G_{LL}(E)=\frac{e^2}{h}\big[1-\operatorname{Re}\det r_L(E)\big].
\label{eq:GLL}
\end{equation}
At $E=0$ and zero temperature, $Q=-1$ exactly when $G_{LL}(0)>e^2/h$, half the quantized value $2e^2/h$ of an isolated Majorana mode~\cite{Law2009}. Without transmission this is the plateau $G_{LL}=(1-Q)e^2/h$ of Ref.~\cite{Wimmer2011}, and simulations had shown $\det r$ falling as the zero-bias peak grows~\cite{DasSarma2016}. In the two-mode model $G_{LL}(0)=(e^2/h)\,2\gamma_L\gamma_R/(\varepsilon^2+\gamma_L\gamma_R)$, which vanishes for a single lead, where Eq.~(\ref{eq:GLL}) reduces to the result of Ref.~\cite{Flensberg2010}. The zero-bias peak splits already at $\varepsilon=\gamma/\sqrt3$, so for $\gamma/\sqrt3<\varepsilon<\gamma$ a split peak coexists with $Q=-1$: the sign of $Q$ follows from the height of the zero-bias conductance, whatever the shape of the peak.

\textit{Bulk transition and Majorana oscillations.} We computed zeros, poles, and scattering matrices of two device models (Sec.~S5~\cite{SM}). The first is a single-subband model of an InAs wire covered with superconducting Al~\cite{Lutchyn2010,Oreg2010}, of length $L=3~\mu$m, with wide-band leads at both ends. A magnetic field $B$ along the wire closes and reopens its bulk gap at $B_c=0.864$~T, above which an infinite wire is topological. The second is a gate-defined two-dimensional (2D) device simulated with the Kwant package~\cite{Groth2014}, a $3~\mu$m$\times100$~nm strip with induced superconductivity, tunnel barriers formed by gates, and normal leads that carry several modes. Equation~(\ref{eq:count}) holds at all 384 settings of the wire and 741 of the 2D device that we tested~\cite{SM}.

In both devices [Figs.~\ref{fig2}(a) and \ref{fig2}(b)] $Q=+1$ at all computed fields below the bulk transition, the first sign change of $Q$, at $B_1=0.8619$~T in the wire, $1.9$~mT below $B_c$, and at $0.8611$~T in the 2D device. A real pair of zeros arrives at $E=0$, coalesces, and splits along the imaginary axis, as the symmetry argument requires. At $B_1$ the largest transmission eigenvalue $T_1$, the transmission probability of the most open channel, equals one at $E=0$ to within $10^{-12}$, and $1-T_1$ grows as $E^4$, with a fitted exponent of $3.99$ in both devices [Fig.~\ref{fig2}(d)]. The bulk transition of a finite mirror-symmetric device is thus an EP of reflectionless states, the perfectly transmitted mode of Ref.~\cite{Akhmerov2011} with a line shape fixed by the double zero.

Above $B_c$ the overlap of the two end modes makes $\varepsilon$ oscillate with field, the Majorana oscillations~\cite{DasSarma2012,Albrecht2016}, and $Q$ changes sign each time $\varepsilon$ crosses $\gamma\approx3~\mu$eV, seven more times below $1.9$~T in each device. Each event is an EP of the same kind: at each of the fourteen, two zeros meet at $E=0$ with $\varepsilon/\gamma$ between $0.98$ and $1.00$, and at the first one in each device they split as $|B-B_{\rm EP}|^{1/2}$~\cite{SM}. The bulk gap stays open at these fields, which depend on the length of the wire [Fig.~\ref{fig2}(c)]. For $L\geq3~\mu$m the first sign change stays within $2$~mT of $B_c$, whereas the later ones move to higher fields as $L$ grows, and for $L\geq5.5~\mu$m none remains below $2$~T.

\textit{Unequal couplings and disorder.} Mirror symmetry is what merges the sign change with the EP. For $\gamma_L\neq\gamma_R$, Eq.~(\ref{eq:twomodes}) keeps both zeros on the imaginary axis until $\varepsilon=(\gamma_L+\gamma_R)/2$, where they coalesce at $i(\gamma_L-\gamma_R)/2$ and leave the axis. The sign change occurs earlier, at $\varepsilon=(\gamma_L\gamma_R)^{1/2}$, when the lower zero passes through $E=0$ alone, so the EP lies on the $Q=+1$ side [Fig.~\ref{fig3}(a)]. Both devices follow these laws when one barrier is made more opaque: in the wire with $\gamma_R/\gamma_L=0.41$ the zeros coalesce $4.1$~mT above the sign change~\cite{SM}. Disorder acts in the same way: in 48 disordered wires with equal barriers the two fields separate by up to $35$~mT, and retuning one barrier restores the double zero at $E=0$ in 38 of them~\cite{SM}.

\textit{Conductance threshold in devices.} Equation~(\ref{eq:GLL}) holds in both devices, whose leads carry both spins and, in the 2D device, several modes. Above the bulk transition the zero-bias conductance and $(e^2/h)[1-\det r_L(0)]$ agree within $0.005\,e^2/h$, at all 741 settings of the 2D device $G_{LL}(0)>e^2/h$ exactly where $Q=-1$~\cite{SM}, and the line shapes of the wire follow the two-mode form within $0.02\,e^2/h$ [Fig.~\ref{fig3}(b)]. The threshold assumes that one Majorana mode carries the coupling to the lead and that the thermal energy $k_BT$ is below the smallest width. Quasi-Majorana states at a smooth barrier violate the second condition, because the width of their weakly coupled mode is below any realistic $k_BT$: in a wire of $6~\mu$m with such a barrier the zero-bias peak at $20$~mK exceeds $e^2/h$ from $0.675$~T on, whereas $Q=+1$ up to $0.845$~T (Sec.~S4~\cite{SM}). In a two-site Kitaev chain, two quantum dots coupled by electron hopping $t$ and by superconducting pairing $\Delta$~\cite{LeijnseFlensberg2012,Dvir2023,tenHaaf2024}, the splitting is $\varepsilon=|t-\Delta|$. With the model parameters of Ref.~\cite{Dvir2023}, the zero-temperature conductance crosses $e^2/h$ at $|t-\Delta|=\gamma$, where $Q$ changes sign, but at the $45$~mK of the experiment the peak stays below $e^2/h$ [Fig.~\ref{fig3}(c)]. Reading $Q$ from the zero-bias conductance requires $k_BT\lesssim0.2\gamma$, about $5$~mK for this chain~\cite{SM}.

\textit{Discussion.} A conductance peak locates a pole, which can sit at zero energy in trivial and topological wires alike~\cite{SanJose2016,Avila2019,Vuik2019}. The zeros tell the two cases apart: for a state coupled to one lead only, every zero is the complex conjugate of a pole and $N_-=0$, whereas a Majorana pair that connects the two leads puts one zero below the real axis. Fits of both local conductance spectra with the two-mode form of Eq.~(\ref{eq:GLL}) return the zeros near $E=0$, and on synthetic spectra of the wire at $20$~mK, where $k_BT=0.57\gamma$, they recover the sign of $Q$ at all 20 fields tested~\cite{SM}. Complex time delays~\cite{Chen2021PRL,Chen2021PRE} and growing or decaying signals (complex-frequency excitation)~\cite{Baranov2017,KimReview2025} give further access to the zeros. A negative $Q$ does not prove a protected phase or set the size of its gap, and quasi-Majorana states can mimic it~\cite{ArayaDay2025}. Equation~(\ref{eq:count}) relies only on particle-hole symmetry and on the positivity of the lead couplings, so it applies equally to multimode wires, longer Kitaev chains~\cite{Bordin2025}, and devices with several leads.

In conclusion, the scattering invariant of a finite superconducting wire is the parity of the number of reflection zeros below the real axis. In mirror-symmetric devices each of its sign changes, including the bulk transition, is an EP of reflectionless states with maximally flat transmission, at which the local conductance crosses $e^2/h$. The zeros of the reflection matrix thus join the poles as measurable markers of the invariant. For Majorana experiments this gives a test of $Q$ that does not rely on the shape of the zero-bias peak and makes clear what $Q=-1$ certifies: a Majorana splitting below the coupling to the leads, not a protected phase.

\textit{Data availability.} The data that support the findings of this Letter are available from the author upon reasonable request.

\bibliography{refs}

\begin{thebibliography}{72}%
\makeatletter
\providecommand \@ifxundefined [1]{%
 \@ifx{#1\undefined}
}%
\providecommand \@ifnum [1]{%
 \ifnum #1\expandafter \@firstoftwo
 \else \expandafter \@secondoftwo
 \fi
}%
\providecommand \@ifx [1]{%
 \ifx #1\expandafter \@firstoftwo
 \else \expandafter \@secondoftwo
 \fi
}%
\providecommand \natexlab [1]{#1}%
\providecommand \enquote  [1]{``#1''}%
\providecommand \bibnamefont  [1]{#1}%
\providecommand \bibfnamefont [1]{#1}%
\providecommand \citenamefont [1]{#1}%
\providecommand \href@noop [0]{\@secondoftwo}%
\providecommand \href [0]{\begingroup \@sanitize@url \@href}%
\providecommand \@href[1]{\@@startlink{#1}\@@href}%
\providecommand \@@href[1]{\endgroup#1\@@endlink}%
\providecommand \@sanitize@url [0]{\catcode `\\12\catcode `\$12\catcode
  `\&12\catcode `\#12\catcode `\^12\catcode `\_12\catcode `\%12\relax}%
\providecommand \@@startlink[1]{}%
\providecommand \@@endlink[0]{}%
\providecommand \url  [0]{\begingroup\@sanitize@url \@url }%
\providecommand \@url [1]{\endgroup\@href {#1}{\urlprefix }}%
\providecommand \urlprefix  [0]{URL }%
\providecommand \Eprint [0]{\href }%
\providecommand \doibase [0]{https://doi.org/}%
\providecommand \selectlanguage [0]{\@gobble}%
\providecommand \bibinfo  [0]{\@secondoftwo}%
\providecommand \bibfield  [0]{\@secondoftwo}%
\providecommand \translation [1]{[#1]}%
\providecommand \BibitemOpen [0]{}%
\providecommand \bibitemStop [0]{}%
\providecommand \bibitemNoStop [0]{.\EOS\space}%
\providecommand \EOS [0]{\spacefactor3000\relax}%
\providecommand \BibitemShut  [1]{\csname bibitem#1\endcsname}%
\let\auto@bib@innerbib\@empty
\bibitem [{\citenamefont {Read}\ and\ \citenamefont
  {Green}(2000)}]{ReadGreen2000}%
  \BibitemOpen
  \bibfield  {author} {\bibinfo {author} {\bibfnamefont {N.}~\bibnamefont
  {Read}}\ and\ \bibinfo {author} {\bibfnamefont {D.}~\bibnamefont {Green}},\
  }\bibfield  {title} {\bibinfo {title} {Paired states of fermions in two
  dimensions with breaking of parity and time-reversal symmetries and the
  fractional quantum {Hall} effect},\ }\href
  {https://doi.org/10.1103/PhysRevB.61.10267} {\bibfield  {journal} {\bibinfo
  {journal} {Physical Review B}\ }\textbf {\bibinfo {volume} {61}},\ \bibinfo
  {pages} {10267} (\bibinfo {year} {2000})}\BibitemShut {NoStop}%
\bibitem [{\citenamefont {Kitaev}(2001)}]{Kitaev2001}%
  \BibitemOpen
  \bibfield  {author} {\bibinfo {author} {\bibfnamefont {A.~Y.}\ \bibnamefont
  {Kitaev}},\ }\bibfield  {title} {\bibinfo {title} {Unpaired {Majorana}
  fermions in quantum wires},\ }\href
  {https://doi.org/10.1070/1063-7869/44/10S/S29} {\bibfield  {journal}
  {\bibinfo  {journal} {Physics-Uspekhi}\ }\textbf {\bibinfo {volume} {44}},\
  \bibinfo {pages} {131} (\bibinfo {year} {2001})}\BibitemShut {NoStop}%
\bibitem [{\citenamefont {Nayak}\ \emph {et~al.}(2008)\citenamefont {Nayak},
  \citenamefont {Simon}, \citenamefont {Stern}, \citenamefont {Freedman},\ and\
  \citenamefont {Das~Sarma}}]{Nayak2008}%
  \BibitemOpen
  \bibfield  {author} {\bibinfo {author} {\bibfnamefont {C.}~\bibnamefont
  {Nayak}}, \bibinfo {author} {\bibfnamefont {S.~H.}\ \bibnamefont {Simon}},
  \bibinfo {author} {\bibfnamefont {A.}~\bibnamefont {Stern}}, \bibinfo
  {author} {\bibfnamefont {M.}~\bibnamefont {Freedman}},\ and\ \bibinfo
  {author} {\bibfnamefont {S.}~\bibnamefont {Das~Sarma}},\ }\bibfield  {title}
  {\bibinfo {title} {Non-{Abelian} anyons and topological quantum
  computation},\ }\href {https://doi.org/10.1103/RevModPhys.80.1083} {\bibfield
   {journal} {\bibinfo  {journal} {Reviews of Modern Physics}\ }\textbf
  {\bibinfo {volume} {80}},\ \bibinfo {pages} {1083} (\bibinfo {year}
  {2008})}\BibitemShut {NoStop}%
\bibitem [{\citenamefont {Alicea}(2012)}]{Alicea2012}%
  \BibitemOpen
  \bibfield  {author} {\bibinfo {author} {\bibfnamefont {J.}~\bibnamefont
  {Alicea}},\ }\bibfield  {title} {\bibinfo {title} {New directions in the
  pursuit of {Majorana} fermions in solid state systems},\ }\href
  {https://doi.org/10.1088/0034-4885/75/7/076501} {\bibfield  {journal}
  {\bibinfo  {journal} {Reports on Progress in Physics}\ }\textbf {\bibinfo
  {volume} {75}},\ \bibinfo {eid} {076501} (\bibinfo {year}
  {2012})}\BibitemShut {NoStop}%
\bibitem [{\citenamefont {Fu}\ and\ \citenamefont {Kane}(2008)}]{FuKane2008}%
  \BibitemOpen
  \bibfield  {author} {\bibinfo {author} {\bibfnamefont {L.}~\bibnamefont
  {Fu}}\ and\ \bibinfo {author} {\bibfnamefont {C.~L.}\ \bibnamefont {Kane}},\
  }\bibfield  {title} {\bibinfo {title} {Superconducting proximity effect and
  {Majorana} fermions at the surface of a topological insulator},\ }\href
  {https://doi.org/10.1103/PhysRevLett.100.096407} {\bibfield  {journal}
  {\bibinfo  {journal} {Physical Review Letters}\ }\textbf {\bibinfo {volume}
  {100}},\ \bibinfo {eid} {096407} (\bibinfo {year} {2008})}\BibitemShut
  {NoStop}%
\bibitem [{\citenamefont {Lutchyn}\ \emph {et~al.}(2010)\citenamefont
  {Lutchyn}, \citenamefont {Sau},\ and\ \citenamefont
  {Das~Sarma}}]{Lutchyn2010}%
  \BibitemOpen
  \bibfield  {author} {\bibinfo {author} {\bibfnamefont {R.~M.}\ \bibnamefont
  {Lutchyn}}, \bibinfo {author} {\bibfnamefont {J.~D.}\ \bibnamefont {Sau}},\
  and\ \bibinfo {author} {\bibfnamefont {S.}~\bibnamefont {Das~Sarma}},\
  }\bibfield  {title} {\bibinfo {title} {{Majorana} fermions and a topological
  phase transition in semiconductor-superconductor heterostructures},\ }\href
  {https://doi.org/10.1103/PhysRevLett.105.077001} {\bibfield  {journal}
  {\bibinfo  {journal} {Physical Review Letters}\ }\textbf {\bibinfo {volume}
  {105}},\ \bibinfo {eid} {077001} (\bibinfo {year} {2010})}\BibitemShut
  {NoStop}%
\bibitem [{\citenamefont {Oreg}\ \emph {et~al.}(2010)\citenamefont {Oreg},
  \citenamefont {Refael},\ and\ \citenamefont {von Oppen}}]{Oreg2010}%
  \BibitemOpen
  \bibfield  {author} {\bibinfo {author} {\bibfnamefont {Y.}~\bibnamefont
  {Oreg}}, \bibinfo {author} {\bibfnamefont {G.}~\bibnamefont {Refael}},\ and\
  \bibinfo {author} {\bibfnamefont {F.}~\bibnamefont {von Oppen}},\ }\bibfield
  {title} {\bibinfo {title} {Helical liquids and {Majorana} bound states in
  quantum wires},\ }\href {https://doi.org/10.1103/PhysRevLett.105.177002}
  {\bibfield  {journal} {\bibinfo  {journal} {Physical Review Letters}\
  }\textbf {\bibinfo {volume} {105}},\ \bibinfo {eid} {177002} (\bibinfo {year}
  {2010})}\BibitemShut {NoStop}%
\bibitem [{\citenamefont {Mourik}\ \emph {et~al.}(2012)\citenamefont {Mourik},
  \citenamefont {Zuo}, \citenamefont {Frolov}, \citenamefont {Plissard},
  \citenamefont {Bakkers},\ and\ \citenamefont {Kouwenhoven}}]{Mourik2012}%
  \BibitemOpen
  \bibfield  {author} {\bibinfo {author} {\bibfnamefont {V.}~\bibnamefont
  {Mourik}}, \bibinfo {author} {\bibfnamefont {K.}~\bibnamefont {Zuo}},
  \bibinfo {author} {\bibfnamefont {S.~M.}\ \bibnamefont {Frolov}}, \bibinfo
  {author} {\bibfnamefont {S.~R.}\ \bibnamefont {Plissard}}, \bibinfo {author}
  {\bibfnamefont {E.~P. A.~M.}\ \bibnamefont {Bakkers}},\ and\ \bibinfo
  {author} {\bibfnamefont {L.~P.}\ \bibnamefont {Kouwenhoven}},\ }\bibfield
  {title} {\bibinfo {title} {Signatures of {Majorana} fermions in hybrid
  superconductor-semiconductor nanowire devices},\ }\href
  {https://doi.org/10.1126/science.1222360} {\bibfield  {journal} {\bibinfo
  {journal} {Science}\ }\textbf {\bibinfo {volume} {336}},\ \bibinfo {pages}
  {1003} (\bibinfo {year} {2012})}\BibitemShut {NoStop}%
\bibitem [{\citenamefont {Kells}\ \emph {et~al.}(2012)\citenamefont {Kells},
  \citenamefont {Meidan},\ and\ \citenamefont {Brouwer}}]{Kells2012}%
  \BibitemOpen
  \bibfield  {author} {\bibinfo {author} {\bibfnamefont {G.}~\bibnamefont
  {Kells}}, \bibinfo {author} {\bibfnamefont {D.}~\bibnamefont {Meidan}},\ and\
  \bibinfo {author} {\bibfnamefont {P.~W.}\ \bibnamefont {Brouwer}},\
  }\bibfield  {title} {\bibinfo {title} {Near-zero-energy end states in
  topologically trivial spin-orbit coupled superconducting nanowires with a
  smooth confinement},\ }\href {https://doi.org/10.1103/PhysRevB.86.100503}
  {\bibfield  {journal} {\bibinfo  {journal} {Physical Review B}\ }\textbf
  {\bibinfo {volume} {86}},\ \bibinfo {eid} {100503} (\bibinfo {year}
  {2012})}\BibitemShut {NoStop}%
\bibitem [{\citenamefont {Liu}\ \emph {et~al.}(2017)\citenamefont {Liu},
  \citenamefont {Sau}, \citenamefont {Stanescu},\ and\ \citenamefont
  {Das~Sarma}}]{Liu2017}%
  \BibitemOpen
  \bibfield  {author} {\bibinfo {author} {\bibfnamefont {C.-X.}\ \bibnamefont
  {Liu}}, \bibinfo {author} {\bibfnamefont {J.~D.}\ \bibnamefont {Sau}},
  \bibinfo {author} {\bibfnamefont {T.~D.}\ \bibnamefont {Stanescu}},\ and\
  \bibinfo {author} {\bibfnamefont {S.}~\bibnamefont {Das~Sarma}},\ }\bibfield
  {title} {\bibinfo {title} {Andreev bound states versus {Majorana} bound
  states in quantum dot-nanowire-superconductor hybrid structures: Trivial
  versus topological zero-bias conductance peaks},\ }\href
  {https://doi.org/10.1103/PhysRevB.96.075161} {\bibfield  {journal} {\bibinfo
  {journal} {Physical Review B}\ }\textbf {\bibinfo {volume} {96}},\ \bibinfo
  {eid} {075161} (\bibinfo {year} {2017})}\BibitemShut {NoStop}%
\bibitem [{\citenamefont {Moore}\ \emph {et~al.}(2018)\citenamefont {Moore},
  \citenamefont {Zeng}, \citenamefont {Stanescu},\ and\ \citenamefont
  {Tewari}}]{Moore2018}%
  \BibitemOpen
  \bibfield  {author} {\bibinfo {author} {\bibfnamefont {C.}~\bibnamefont
  {Moore}}, \bibinfo {author} {\bibfnamefont {C.}~\bibnamefont {Zeng}},
  \bibinfo {author} {\bibfnamefont {T.~D.}\ \bibnamefont {Stanescu}},\ and\
  \bibinfo {author} {\bibfnamefont {S.}~\bibnamefont {Tewari}},\ }\bibfield
  {title} {\bibinfo {title} {Quantized zero-bias conductance plateau in
  semiconductor-superconductor heterostructures without topological {M}ajorana
  zero modes},\ }\href {https://doi.org/10.1103/PhysRevB.98.155314} {\bibfield
  {journal} {\bibinfo  {journal} {Physical Review B}\ }\textbf {\bibinfo
  {volume} {98}},\ \bibinfo {eid} {155314} (\bibinfo {year}
  {2018})}\BibitemShut {NoStop}%
\bibitem [{\citenamefont {Vuik}\ \emph {et~al.}(2019)\citenamefont {Vuik},
  \citenamefont {Nijholt}, \citenamefont {Akhmerov},\ and\ \citenamefont
  {Wimmer}}]{Vuik2019}%
  \BibitemOpen
  \bibfield  {author} {\bibinfo {author} {\bibfnamefont {A.}~\bibnamefont
  {Vuik}}, \bibinfo {author} {\bibfnamefont {B.}~\bibnamefont {Nijholt}},
  \bibinfo {author} {\bibfnamefont {A.~R.}\ \bibnamefont {Akhmerov}},\ and\
  \bibinfo {author} {\bibfnamefont {M.}~\bibnamefont {Wimmer}},\ }\bibfield
  {title} {\bibinfo {title} {Reproducing topological properties with
  quasi-{M}ajorana states},\ }\href
  {https://doi.org/10.21468/SciPostPhys.7.5.061} {\bibfield  {journal}
  {\bibinfo  {journal} {SciPost Physics}\ }\textbf {\bibinfo {volume} {7}},\
  \bibinfo {eid} {061} (\bibinfo {year} {2019})}\BibitemShut {NoStop}%
\bibitem [{\citenamefont {Prada}\ \emph {et~al.}(2020)\citenamefont {Prada},
  \citenamefont {San-Jose}, \citenamefont {de~Moor}, \citenamefont {Geresdi},
  \citenamefont {Lee}, \citenamefont {Klinovaja}, \citenamefont {Loss},
  \citenamefont {Nyg{\aa}rd}, \citenamefont {Aguado},\ and\ \citenamefont
  {Kouwenhoven}}]{Prada2020}%
  \BibitemOpen
  \bibfield  {author} {\bibinfo {author} {\bibfnamefont {E.}~\bibnamefont
  {Prada}}, \bibinfo {author} {\bibfnamefont {P.}~\bibnamefont {San-Jose}},
  \bibinfo {author} {\bibfnamefont {M.~W.~A.}\ \bibnamefont {de~Moor}},
  \bibinfo {author} {\bibfnamefont {A.}~\bibnamefont {Geresdi}}, \bibinfo
  {author} {\bibfnamefont {E.~J.~H.}\ \bibnamefont {Lee}}, \bibinfo {author}
  {\bibfnamefont {J.}~\bibnamefont {Klinovaja}}, \bibinfo {author}
  {\bibfnamefont {D.}~\bibnamefont {Loss}}, \bibinfo {author} {\bibfnamefont
  {J.}~\bibnamefont {Nyg{\aa}rd}}, \bibinfo {author} {\bibfnamefont
  {R.}~\bibnamefont {Aguado}},\ and\ \bibinfo {author} {\bibfnamefont {L.~P.}\
  \bibnamefont {Kouwenhoven}},\ }\bibfield  {title} {\bibinfo {title} {From
  {Andreev} to {Majorana} bound states in hybrid superconductor-semiconductor
  nanowires},\ }\href {https://doi.org/10.1038/s42254-020-0228-y} {\bibfield
  {journal} {\bibinfo  {journal} {Nature Reviews Physics}\ }\textbf {\bibinfo
  {volume} {2}},\ \bibinfo {pages} {575} (\bibinfo {year} {2020})}\BibitemShut
  {NoStop}%
\bibitem [{\citenamefont {Pikulin}\ \emph {et~al.}(2021)\citenamefont
  {Pikulin}, \citenamefont {van Heck}, \citenamefont {Karzig}, \citenamefont
  {Martinez}, \citenamefont {Nijholt}, \citenamefont {Laeven}, \citenamefont
  {Winkler}, \citenamefont {Watson}, \citenamefont {Heedt}, \citenamefont
  {Temurhan}, \citenamefont {Svidenko}, \citenamefont {Lutchyn}, \citenamefont
  {Thomas}, \citenamefont {de~Lange}, \citenamefont {Casparis},\ and\
  \citenamefont {Nayak}}]{Pikulin2021}%
  \BibitemOpen
  \bibfield  {author} {\bibinfo {author} {\bibfnamefont {D.~I.}\ \bibnamefont
  {Pikulin}}, \bibinfo {author} {\bibfnamefont {B.}~\bibnamefont {van Heck}},
  \bibinfo {author} {\bibfnamefont {T.}~\bibnamefont {Karzig}}, \bibinfo
  {author} {\bibfnamefont {E.~A.}\ \bibnamefont {Martinez}}, \bibinfo {author}
  {\bibfnamefont {B.}~\bibnamefont {Nijholt}}, \bibinfo {author} {\bibfnamefont
  {T.}~\bibnamefont {Laeven}}, \bibinfo {author} {\bibfnamefont {G.~W.}\
  \bibnamefont {Winkler}}, \bibinfo {author} {\bibfnamefont {J.~D.}\
  \bibnamefont {Watson}}, \bibinfo {author} {\bibfnamefont {S.}~\bibnamefont
  {Heedt}}, \bibinfo {author} {\bibfnamefont {M.}~\bibnamefont {Temurhan}},
  \bibinfo {author} {\bibfnamefont {V.}~\bibnamefont {Svidenko}}, \bibinfo
  {author} {\bibfnamefont {R.~M.}\ \bibnamefont {Lutchyn}}, \bibinfo {author}
  {\bibfnamefont {M.}~\bibnamefont {Thomas}}, \bibinfo {author} {\bibfnamefont
  {G.}~\bibnamefont {de~Lange}}, \bibinfo {author} {\bibfnamefont
  {L.}~\bibnamefont {Casparis}},\ and\ \bibinfo {author} {\bibfnamefont
  {C.}~\bibnamefont {Nayak}},\ }\href@noop {} {\bibinfo {title} {Protocol to
  identify a topological superconducting phase in a three-terminal device}}
  (\bibinfo {year} {2021}),\ \Eprint {https://arxiv.org/abs/2103.12217}
  {arXiv:2103.12217} \BibitemShut {NoStop}%
\bibitem [{\citenamefont {Aghaee}\ \emph {et~al.}(2023)\citenamefont {Aghaee}
  \emph {et~al.}}]{Aghaee2023TGP}%
  \BibitemOpen
  \bibfield  {author} {\bibinfo {author} {\bibfnamefont {M.}~\bibnamefont
  {Aghaee}} \emph {et~al.} (\bibinfo {collaboration} {Microsoft Quantum}),\
  }\bibfield  {title} {\bibinfo {title} {{InAs-Al} hybrid devices passing the
  topological gap protocol},\ }\href
  {https://doi.org/10.1103/PhysRevB.107.245423} {\bibfield  {journal} {\bibinfo
   {journal} {Physical Review B}\ }\textbf {\bibinfo {volume} {107}},\ \bibinfo
  {eid} {245423} (\bibinfo {year} {2023})}\BibitemShut {NoStop}%
\bibitem [{\citenamefont {{Microsoft Azure Quantum}}\ \emph
  {et~al.}(2025)\citenamefont {{Microsoft Azure Quantum}}, \citenamefont
  {Aghaee} \emph {et~al.}}]{Microsoft2025Nature}%
  \BibitemOpen
  \bibfield  {author} {\bibinfo {author} {\bibnamefont {{Microsoft Azure
  Quantum}}}, \bibinfo {author} {\bibfnamefont {M.}~\bibnamefont {Aghaee}},
  \emph {et~al.},\ }\bibfield  {title} {\bibinfo {title} {Interferometric
  single-shot parity measurement in {InAs--Al} hybrid devices},\ }\href
  {https://doi.org/10.1038/s41586-024-08445-2} {\bibfield  {journal} {\bibinfo
  {journal} {Nature}\ }\textbf {\bibinfo {volume} {638}},\ \bibinfo {pages}
  {651} (\bibinfo {year} {2025})}\BibitemShut {NoStop}%
\bibitem [{\citenamefont {Legg}(2025)}]{Legg2025TGP}%
  \BibitemOpen
  \bibfield  {author} {\bibinfo {author} {\bibfnamefont {H.~F.}\ \bibnamefont
  {Legg}},\ }\href@noop {} {\bibinfo {title} {Comment on ``{InAs-Al} hybrid
  devices passing the topological gap protocol''}} (\bibinfo {year} {2025}),\
  \Eprint {https://arxiv.org/abs/2502.19560} {arXiv:2502.19560} \BibitemShut
  {NoStop}%
\bibitem [{\citenamefont {Legg}(2026)}]{Legg2026Nature}%
  \BibitemOpen
  \bibfield  {author} {\bibinfo {author} {\bibfnamefont {H.~F.}\ \bibnamefont
  {Legg}},\ }\bibfield  {title} {\bibinfo {title} {On the robustness of
  topological gap detection via transport},\ }\href
  {https://doi.org/10.1038/s41586-026-10567-8} {\bibfield  {journal} {\bibinfo
  {journal} {Nature}\ }\textbf {\bibinfo {volume} {654}},\ \bibinfo {pages}
  {E22} (\bibinfo {year} {2026})}\BibitemShut {NoStop}%
\bibitem [{\citenamefont {Aghaee}\ \emph {et~al.}(2025)\citenamefont {Aghaee}
  \emph {et~al.}}]{Aghaee2025Response}%
  \BibitemOpen
  \bibfield  {author} {\bibinfo {author} {\bibfnamefont {M.}~\bibnamefont
  {Aghaee}} \emph {et~al.},\ }\href@noop {} {\bibinfo {title} {Response to
  recent comments on {Phys. Rev. B} 107, 245423 (2023) and {Subsection S4.3} of
  the {Supp. Info.} for {Nature} 638, 651-655 (2025)}} (\bibinfo {year}
  {2025}),\ \Eprint {https://arxiv.org/abs/2504.13240} {arXiv:2504.13240}
  \BibitemShut {NoStop}%
\bibitem [{\citenamefont {Leijnse}\ and\ \citenamefont
  {Flensberg}(2012)}]{LeijnseFlensberg2012}%
  \BibitemOpen
  \bibfield  {author} {\bibinfo {author} {\bibfnamefont {M.}~\bibnamefont
  {Leijnse}}\ and\ \bibinfo {author} {\bibfnamefont {K.}~\bibnamefont
  {Flensberg}},\ }\bibfield  {title} {\bibinfo {title} {Parity qubits and poor
  man's {Majorana} bound states in double quantum dots},\ }\href
  {https://doi.org/10.1103/PhysRevB.86.134528} {\bibfield  {journal} {\bibinfo
  {journal} {Physical Review B}\ }\textbf {\bibinfo {volume} {86}},\ \bibinfo
  {eid} {134528} (\bibinfo {year} {2012})}\BibitemShut {NoStop}%
\bibitem [{\citenamefont {Dvir}\ \emph {et~al.}(2023)\citenamefont {Dvir},
  \citenamefont {Wang}, \citenamefont {van Loo}, \citenamefont {Liu},
  \citenamefont {Mazur}, \citenamefont {Bordin}, \citenamefont {ten Haaf},
  \citenamefont {Wang}, \citenamefont {van Driel}, \citenamefont {Zatelli},
  \citenamefont {Li}, \citenamefont {Malinowski}, \citenamefont {Gazibegovic},
  \citenamefont {Badawy}, \citenamefont {Bakkers}, \citenamefont {Wimmer},\
  and\ \citenamefont {Kouwenhoven}}]{Dvir2023}%
  \BibitemOpen
  \bibfield  {author} {\bibinfo {author} {\bibfnamefont {T.}~\bibnamefont
  {Dvir}}, \bibinfo {author} {\bibfnamefont {G.}~\bibnamefont {Wang}}, \bibinfo
  {author} {\bibfnamefont {N.}~\bibnamefont {van Loo}}, \bibinfo {author}
  {\bibfnamefont {C.-X.}\ \bibnamefont {Liu}}, \bibinfo {author} {\bibfnamefont
  {G.~P.}\ \bibnamefont {Mazur}}, \bibinfo {author} {\bibfnamefont
  {A.}~\bibnamefont {Bordin}}, \bibinfo {author} {\bibfnamefont {S.~L.~D.}\
  \bibnamefont {ten Haaf}}, \bibinfo {author} {\bibfnamefont {J.-Y.}\
  \bibnamefont {Wang}}, \bibinfo {author} {\bibfnamefont {D.}~\bibnamefont {van
  Driel}}, \bibinfo {author} {\bibfnamefont {F.}~\bibnamefont {Zatelli}},
  \bibinfo {author} {\bibfnamefont {X.}~\bibnamefont {Li}}, \bibinfo {author}
  {\bibfnamefont {F.~K.}\ \bibnamefont {Malinowski}}, \bibinfo {author}
  {\bibfnamefont {S.}~\bibnamefont {Gazibegovic}}, \bibinfo {author}
  {\bibfnamefont {G.}~\bibnamefont {Badawy}}, \bibinfo {author} {\bibfnamefont
  {E.~P. A.~M.}\ \bibnamefont {Bakkers}}, \bibinfo {author} {\bibfnamefont
  {M.}~\bibnamefont {Wimmer}},\ and\ \bibinfo {author} {\bibfnamefont {L.~P.}\
  \bibnamefont {Kouwenhoven}},\ }\bibfield  {title} {\bibinfo {title}
  {Realization of a minimal {Kitaev} chain in coupled quantum dots},\ }\href
  {https://doi.org/10.1038/s41586-022-05585-1} {\bibfield  {journal} {\bibinfo
  {journal} {Nature}\ }\textbf {\bibinfo {volume} {614}},\ \bibinfo {pages}
  {445} (\bibinfo {year} {2023})}\BibitemShut {NoStop}%
\bibitem [{\citenamefont {ten Haaf}\ \emph {et~al.}(2024)\citenamefont {ten
  Haaf}, \citenamefont {Wang}, \citenamefont {Bozkurt}, \citenamefont {Liu},
  \citenamefont {Kulesh}, \citenamefont {Kim}, \citenamefont {Xiao},
  \citenamefont {Thomas}, \citenamefont {Manfra}, \citenamefont {Dvir},
  \citenamefont {Wimmer},\ and\ \citenamefont {Goswami}}]{tenHaaf2024}%
  \BibitemOpen
  \bibfield  {author} {\bibinfo {author} {\bibfnamefont {S.~L.~D.}\
  \bibnamefont {ten Haaf}}, \bibinfo {author} {\bibfnamefont {Q.}~\bibnamefont
  {Wang}}, \bibinfo {author} {\bibfnamefont {A.~M.}\ \bibnamefont {Bozkurt}},
  \bibinfo {author} {\bibfnamefont {C.-X.}\ \bibnamefont {Liu}}, \bibinfo
  {author} {\bibfnamefont {I.}~\bibnamefont {Kulesh}}, \bibinfo {author}
  {\bibfnamefont {P.}~\bibnamefont {Kim}}, \bibinfo {author} {\bibfnamefont
  {D.}~\bibnamefont {Xiao}}, \bibinfo {author} {\bibfnamefont {C.}~\bibnamefont
  {Thomas}}, \bibinfo {author} {\bibfnamefont {M.~J.}\ \bibnamefont {Manfra}},
  \bibinfo {author} {\bibfnamefont {T.}~\bibnamefont {Dvir}}, \bibinfo {author}
  {\bibfnamefont {M.}~\bibnamefont {Wimmer}},\ and\ \bibinfo {author}
  {\bibfnamefont {S.}~\bibnamefont {Goswami}},\ }\bibfield  {title} {\bibinfo
  {title} {A two-site {K}itaev chain in a two-dimensional electron gas},\
  }\href {https://doi.org/10.1038/s41586-024-07434-9} {\bibfield  {journal}
  {\bibinfo  {journal} {Nature}\ }\textbf {\bibinfo {volume} {630}},\ \bibinfo
  {pages} {329} (\bibinfo {year} {2024})}\BibitemShut {NoStop}%
\bibitem [{\citenamefont {Bordin}\ \emph {et~al.}(2025)\citenamefont {Bordin},
  \citenamefont {Liu}, \citenamefont {Dvir}, \citenamefont {Zatelli},
  \citenamefont {ten Haaf}, \citenamefont {van Driel}, \citenamefont {Wang},
  \citenamefont {van Loo}, \citenamefont {Zhang}, \citenamefont {Wolff},
  \citenamefont {Van~Caekenberghe}, \citenamefont {Badawy}, \citenamefont
  {Gazibegovic}, \citenamefont {Bakkers}, \citenamefont {Wimmer}, \citenamefont
  {Kouwenhoven},\ and\ \citenamefont {Mazur}}]{Bordin2025}%
  \BibitemOpen
  \bibfield  {author} {\bibinfo {author} {\bibfnamefont {A.}~\bibnamefont
  {Bordin}}, \bibinfo {author} {\bibfnamefont {C.-X.}\ \bibnamefont {Liu}},
  \bibinfo {author} {\bibfnamefont {T.}~\bibnamefont {Dvir}}, \bibinfo {author}
  {\bibfnamefont {F.}~\bibnamefont {Zatelli}}, \bibinfo {author} {\bibfnamefont
  {S.~L.~D.}\ \bibnamefont {ten Haaf}}, \bibinfo {author} {\bibfnamefont
  {D.}~\bibnamefont {van Driel}}, \bibinfo {author} {\bibfnamefont
  {G.}~\bibnamefont {Wang}}, \bibinfo {author} {\bibfnamefont {N.}~\bibnamefont
  {van Loo}}, \bibinfo {author} {\bibfnamefont {Y.}~\bibnamefont {Zhang}},
  \bibinfo {author} {\bibfnamefont {J.~C.}\ \bibnamefont {Wolff}}, \bibinfo
  {author} {\bibfnamefont {T.}~\bibnamefont {Van~Caekenberghe}}, \bibinfo
  {author} {\bibfnamefont {G.}~\bibnamefont {Badawy}}, \bibinfo {author}
  {\bibfnamefont {S.}~\bibnamefont {Gazibegovic}}, \bibinfo {author}
  {\bibfnamefont {E.~P. A.~M.}\ \bibnamefont {Bakkers}}, \bibinfo {author}
  {\bibfnamefont {M.}~\bibnamefont {Wimmer}}, \bibinfo {author} {\bibfnamefont
  {L.~P.}\ \bibnamefont {Kouwenhoven}},\ and\ \bibinfo {author} {\bibfnamefont
  {G.~P.}\ \bibnamefont {Mazur}},\ }\bibfield  {title} {\bibinfo {title}
  {Enhanced {M}ajorana stability in a three-site {K}itaev chain},\ }\href
  {https://doi.org/10.1038/s41565-025-01894-4} {\bibfield  {journal} {\bibinfo
  {journal} {Nature Nanotechnology}\ }\textbf {\bibinfo {volume} {20}},\
  \bibinfo {pages} {726} (\bibinfo {year} {2025})}\BibitemShut {NoStop}%
\bibitem [{\citenamefont {Akhmerov}\ \emph {et~al.}(2011)\citenamefont
  {Akhmerov}, \citenamefont {Dahlhaus}, \citenamefont {Hassler}, \citenamefont
  {Wimmer},\ and\ \citenamefont {Beenakker}}]{Akhmerov2011}%
  \BibitemOpen
  \bibfield  {author} {\bibinfo {author} {\bibfnamefont {A.~R.}\ \bibnamefont
  {Akhmerov}}, \bibinfo {author} {\bibfnamefont {J.~P.}\ \bibnamefont
  {Dahlhaus}}, \bibinfo {author} {\bibfnamefont {F.}~\bibnamefont {Hassler}},
  \bibinfo {author} {\bibfnamefont {M.}~\bibnamefont {Wimmer}},\ and\ \bibinfo
  {author} {\bibfnamefont {C.~W.~J.}\ \bibnamefont {Beenakker}},\ }\bibfield
  {title} {\bibinfo {title} {Quantized conductance at the {M}ajorana phase
  transition in a disordered superconducting wire},\ }\href
  {https://doi.org/10.1103/PhysRevLett.106.057001} {\bibfield  {journal}
  {\bibinfo  {journal} {Physical Review Letters}\ }\textbf {\bibinfo {volume}
  {106}},\ \bibinfo {eid} {057001} (\bibinfo {year} {2011})}\BibitemShut
  {NoStop}%
\bibitem [{\citenamefont {Fulga}\ \emph {et~al.}(2011)\citenamefont {Fulga},
  \citenamefont {Hassler}, \citenamefont {Akhmerov},\ and\ \citenamefont
  {Beenakker}}]{Fulga2011}%
  \BibitemOpen
  \bibfield  {author} {\bibinfo {author} {\bibfnamefont {I.~C.}\ \bibnamefont
  {Fulga}}, \bibinfo {author} {\bibfnamefont {F.}~\bibnamefont {Hassler}},
  \bibinfo {author} {\bibfnamefont {A.~R.}\ \bibnamefont {Akhmerov}},\ and\
  \bibinfo {author} {\bibfnamefont {C.~W.~J.}\ \bibnamefont {Beenakker}},\
  }\bibfield  {title} {\bibinfo {title} {Scattering formula for the topological
  quantum number of a disordered multimode wire},\ }\href
  {https://doi.org/10.1103/PhysRevB.83.155429} {\bibfield  {journal} {\bibinfo
  {journal} {Physical Review B}\ }\textbf {\bibinfo {volume} {83}},\ \bibinfo
  {eid} {155429} (\bibinfo {year} {2011})}\BibitemShut {NoStop}%
\bibitem [{\citenamefont {Fulga}\ \emph {et~al.}(2012)\citenamefont {Fulga},
  \citenamefont {Hassler},\ and\ \citenamefont {Akhmerov}}]{Fulga2012}%
  \BibitemOpen
  \bibfield  {author} {\bibinfo {author} {\bibfnamefont {I.~C.}\ \bibnamefont
  {Fulga}}, \bibinfo {author} {\bibfnamefont {F.}~\bibnamefont {Hassler}},\
  and\ \bibinfo {author} {\bibfnamefont {A.~R.}\ \bibnamefont {Akhmerov}},\
  }\bibfield  {title} {\bibinfo {title} {Scattering theory of topological
  insulators and superconductors},\ }\href
  {https://doi.org/10.1103/PhysRevB.85.165409} {\bibfield  {journal} {\bibinfo
  {journal} {Physical Review B}\ }\textbf {\bibinfo {volume} {85}},\ \bibinfo
  {eid} {165409} (\bibinfo {year} {2012})}\BibitemShut {NoStop}%
\bibitem [{\citenamefont {Beenakker}(2015)}]{Beenakker2015}%
  \BibitemOpen
  \bibfield  {author} {\bibinfo {author} {\bibfnamefont {C.~W.~J.}\
  \bibnamefont {Beenakker}},\ }\bibfield  {title} {\bibinfo {title}
  {Random-matrix theory of {Majorana} fermions and topological
  superconductors},\ }\href {https://doi.org/10.1103/RevModPhys.87.1037}
  {\bibfield  {journal} {\bibinfo  {journal} {Reviews of Modern Physics}\
  }\textbf {\bibinfo {volume} {87}},\ \bibinfo {pages} {1037} (\bibinfo {year}
  {2015})}\BibitemShut {NoStop}%
\bibitem [{\citenamefont {Pikulin}\ and\ \citenamefont
  {Nazarov}(2012)}]{Pikulin2012}%
  \BibitemOpen
  \bibfield  {author} {\bibinfo {author} {\bibfnamefont {D.~I.}\ \bibnamefont
  {Pikulin}}\ and\ \bibinfo {author} {\bibfnamefont {Y.~V.}\ \bibnamefont
  {Nazarov}},\ }\bibfield  {title} {\bibinfo {title} {Topological properties of
  superconducting junctions},\ }\href
  {https://doi.org/10.1134/S0021364011210090} {\bibfield  {journal} {\bibinfo
  {journal} {JETP Letters}\ }\textbf {\bibinfo {volume} {94}},\ \bibinfo
  {pages} {693} (\bibinfo {year} {2012})}\BibitemShut {NoStop}%
\bibitem [{\citenamefont {Araya~Day}\ \emph {et~al.}(2025)\citenamefont
  {Araya~Day}, \citenamefont {Manesco}, \citenamefont {Wimmer},\ and\
  \citenamefont {Akhmerov}}]{ArayaDay2025}%
  \BibitemOpen
  \bibfield  {author} {\bibinfo {author} {\bibfnamefont {I.}~\bibnamefont
  {Araya~Day}}, \bibinfo {author} {\bibfnamefont {A.~L.~R.}\ \bibnamefont
  {Manesco}}, \bibinfo {author} {\bibfnamefont {M.}~\bibnamefont {Wimmer}},\
  and\ \bibinfo {author} {\bibfnamefont {A.~R.}\ \bibnamefont {Akhmerov}},\
  }\bibfield  {title} {\bibinfo {title} {Identifying biases of the {M}ajorana
  scattering invariant},\ }\href
  {https://doi.org/10.21468/SciPostPhysCore.8.3.047} {\bibfield  {journal}
  {\bibinfo  {journal} {SciPost Physics Core}\ }\textbf {\bibinfo {volume}
  {8}},\ \bibinfo {eid} {047} (\bibinfo {year} {2025})}\BibitemShut {NoStop}%
\bibitem [{\citenamefont {Law}\ \emph {et~al.}(2009)\citenamefont {Law},
  \citenamefont {Lee},\ and\ \citenamefont {Ng}}]{Law2009}%
  \BibitemOpen
  \bibfield  {author} {\bibinfo {author} {\bibfnamefont {K.~T.}\ \bibnamefont
  {Law}}, \bibinfo {author} {\bibfnamefont {P.~A.}\ \bibnamefont {Lee}},\ and\
  \bibinfo {author} {\bibfnamefont {T.~K.}\ \bibnamefont {Ng}},\ }\bibfield
  {title} {\bibinfo {title} {{M}ajorana fermion induced resonant {A}ndreev
  reflection},\ }\href {https://doi.org/10.1103/PhysRevLett.103.237001}
  {\bibfield  {journal} {\bibinfo  {journal} {Physical Review Letters}\
  }\textbf {\bibinfo {volume} {103}},\ \bibinfo {eid} {237001} (\bibinfo {year}
  {2009})}\BibitemShut {NoStop}%
\bibitem [{\citenamefont {Flensberg}(2010)}]{Flensberg2010}%
  \BibitemOpen
  \bibfield  {author} {\bibinfo {author} {\bibfnamefont {K.}~\bibnamefont
  {Flensberg}},\ }\bibfield  {title} {\bibinfo {title} {Tunneling
  characteristics of a chain of {M}ajorana bound states},\ }\href
  {https://doi.org/10.1103/PhysRevB.82.180516} {\bibfield  {journal} {\bibinfo
  {journal} {Physical Review B}\ }\textbf {\bibinfo {volume} {82}},\ \bibinfo
  {eid} {180516} (\bibinfo {year} {2010})}\BibitemShut {NoStop}%
\bibitem [{\citenamefont {Pikulin}\ and\ \citenamefont
  {Nazarov}(2013)}]{Pikulin2013}%
  \BibitemOpen
  \bibfield  {author} {\bibinfo {author} {\bibfnamefont {D.~I.}\ \bibnamefont
  {Pikulin}}\ and\ \bibinfo {author} {\bibfnamefont {Y.~V.}\ \bibnamefont
  {Nazarov}},\ }\bibfield  {title} {\bibinfo {title} {Two types of topological
  transitions in finite {M}ajorana wires},\ }\href
  {https://doi.org/10.1103/PhysRevB.87.235421} {\bibfield  {journal} {\bibinfo
  {journal} {Physical Review B}\ }\textbf {\bibinfo {volume} {87}},\ \bibinfo
  {eid} {235421} (\bibinfo {year} {2013})}\BibitemShut {NoStop}%
\bibitem [{\citenamefont {Heiss}(2012)}]{Heiss2012}%
  \BibitemOpen
  \bibfield  {author} {\bibinfo {author} {\bibfnamefont {W.~D.}\ \bibnamefont
  {Heiss}},\ }\bibfield  {title} {\bibinfo {title} {The physics of exceptional
  points},\ }\href {https://doi.org/10.1088/1751-8113/45/44/444016} {\bibfield
  {journal} {\bibinfo  {journal} {Journal of Physics A: Mathematical and
  Theoretical}\ }\textbf {\bibinfo {volume} {45}},\ \bibinfo {eid} {444016}
  (\bibinfo {year} {2012})}\BibitemShut {NoStop}%
\bibitem [{\citenamefont {Miri}\ and\ \citenamefont
  {Al{\`u}}(2019)}]{MiriAlu2019}%
  \BibitemOpen
  \bibfield  {author} {\bibinfo {author} {\bibfnamefont {M.-A.}\ \bibnamefont
  {Miri}}\ and\ \bibinfo {author} {\bibfnamefont {A.}~\bibnamefont {Al{\`u}}},\
  }\bibfield  {title} {\bibinfo {title} {Exceptional points in optics and
  photonics},\ }\href {https://doi.org/10.1126/science.aar7709} {\bibfield
  {journal} {\bibinfo  {journal} {Science}\ }\textbf {\bibinfo {volume}
  {363}},\ \bibinfo {eid} {eaar7709} (\bibinfo {year} {2019})}\BibitemShut
  {NoStop}%
\bibitem [{\citenamefont {Bergholtz}\ \emph {et~al.}(2021)\citenamefont
  {Bergholtz}, \citenamefont {Budich},\ and\ \citenamefont
  {Kunst}}]{Bergholtz2021}%
  \BibitemOpen
  \bibfield  {author} {\bibinfo {author} {\bibfnamefont {E.~J.}\ \bibnamefont
  {Bergholtz}}, \bibinfo {author} {\bibfnamefont {J.~C.}\ \bibnamefont
  {Budich}},\ and\ \bibinfo {author} {\bibfnamefont {F.~K.}\ \bibnamefont
  {Kunst}},\ }\bibfield  {title} {\bibinfo {title} {Exceptional topology of
  non-{Hermitian} systems},\ }\href
  {https://doi.org/10.1103/RevModPhys.93.015005} {\bibfield  {journal}
  {\bibinfo  {journal} {Reviews of Modern Physics}\ }\textbf {\bibinfo {volume}
  {93}},\ \bibinfo {eid} {015005} (\bibinfo {year} {2021})}\BibitemShut
  {NoStop}%
\bibitem [{\citenamefont {Ashida}\ \emph {et~al.}(2020)\citenamefont {Ashida},
  \citenamefont {Gong},\ and\ \citenamefont {Ueda}}]{Ashida2020}%
  \BibitemOpen
  \bibfield  {author} {\bibinfo {author} {\bibfnamefont {Y.}~\bibnamefont
  {Ashida}}, \bibinfo {author} {\bibfnamefont {Z.}~\bibnamefont {Gong}},\ and\
  \bibinfo {author} {\bibfnamefont {M.}~\bibnamefont {Ueda}},\ }\bibfield
  {title} {\bibinfo {title} {Non-{Hermitian} physics},\ }\href
  {https://doi.org/10.1080/00018732.2021.1876991} {\bibfield  {journal}
  {\bibinfo  {journal} {Advances in Physics}\ }\textbf {\bibinfo {volume}
  {69}},\ \bibinfo {pages} {249} (\bibinfo {year} {2020})}\BibitemShut
  {NoStop}%
\bibitem [{\citenamefont {San-Jose}\ \emph {et~al.}(2016)\citenamefont
  {San-Jose}, \citenamefont {Cayao}, \citenamefont {Prada},\ and\ \citenamefont
  {Aguado}}]{SanJose2016}%
  \BibitemOpen
  \bibfield  {author} {\bibinfo {author} {\bibfnamefont {P.}~\bibnamefont
  {San-Jose}}, \bibinfo {author} {\bibfnamefont {J.}~\bibnamefont {Cayao}},
  \bibinfo {author} {\bibfnamefont {E.}~\bibnamefont {Prada}},\ and\ \bibinfo
  {author} {\bibfnamefont {R.}~\bibnamefont {Aguado}},\ }\bibfield  {title}
  {\bibinfo {title} {Majorana bound states from exceptional points in
  non-topological superconductors},\ }\href {https://doi.org/10.1038/srep21427}
  {\bibfield  {journal} {\bibinfo  {journal} {Scientific Reports}\ }\textbf
  {\bibinfo {volume} {6}},\ \bibinfo {eid} {21427} (\bibinfo {year}
  {2016})}\BibitemShut {NoStop}%
\bibitem [{\citenamefont {Avila}\ \emph {et~al.}(2019)\citenamefont {Avila},
  \citenamefont {Pe{\~n}aranda}, \citenamefont {Prada}, \citenamefont
  {San-Jose},\ and\ \citenamefont {Aguado}}]{Avila2019}%
  \BibitemOpen
  \bibfield  {author} {\bibinfo {author} {\bibfnamefont {J.}~\bibnamefont
  {Avila}}, \bibinfo {author} {\bibfnamefont {F.}~\bibnamefont
  {Pe{\~n}aranda}}, \bibinfo {author} {\bibfnamefont {E.}~\bibnamefont
  {Prada}}, \bibinfo {author} {\bibfnamefont {P.}~\bibnamefont {San-Jose}},\
  and\ \bibinfo {author} {\bibfnamefont {R.}~\bibnamefont {Aguado}},\
  }\bibfield  {title} {\bibinfo {title} {Non-{H}ermitian topology as a unifying
  framework for the {A}ndreev versus {M}ajorana states controversy},\ }\href
  {https://doi.org/10.1038/s42005-019-0231-8} {\bibfield  {journal} {\bibinfo
  {journal} {Communications Physics}\ }\textbf {\bibinfo {volume} {2}},\
  \bibinfo {eid} {133} (\bibinfo {year} {2019})}\BibitemShut {NoStop}%
\bibitem [{\citenamefont {Grigoriev}\ \emph {et~al.}(2013)\citenamefont
  {Grigoriev}, \citenamefont {Tahri}, \citenamefont {Varault}, \citenamefont
  {Rolly}, \citenamefont {Stout}, \citenamefont {Wenger},\ and\ \citenamefont
  {Bonod}}]{Grigoriev2013}%
  \BibitemOpen
  \bibfield  {author} {\bibinfo {author} {\bibfnamefont {V.}~\bibnamefont
  {Grigoriev}}, \bibinfo {author} {\bibfnamefont {A.}~\bibnamefont {Tahri}},
  \bibinfo {author} {\bibfnamefont {S.}~\bibnamefont {Varault}}, \bibinfo
  {author} {\bibfnamefont {B.}~\bibnamefont {Rolly}}, \bibinfo {author}
  {\bibfnamefont {B.}~\bibnamefont {Stout}}, \bibinfo {author} {\bibfnamefont
  {J.}~\bibnamefont {Wenger}},\ and\ \bibinfo {author} {\bibfnamefont
  {N.}~\bibnamefont {Bonod}},\ }\bibfield  {title} {\bibinfo {title}
  {Optimization of resonant effects in nanostructures via {Weierstrass}
  factorization},\ }\href {https://doi.org/10.1103/PhysRevA.88.011803}
  {\bibfield  {journal} {\bibinfo  {journal} {Physical Review A}\ }\textbf
  {\bibinfo {volume} {88}},\ \bibinfo {eid} {011803} (\bibinfo {year}
  {2013})}\BibitemShut {NoStop}%
\bibitem [{\citenamefont {Krasnok}\ \emph {et~al.}(2019)\citenamefont
  {Krasnok}, \citenamefont {Baranov}, \citenamefont {Li}, \citenamefont {Miri},
  \citenamefont {Monticone},\ and\ \citenamefont {Al{\`u}}}]{Krasnok2019}%
  \BibitemOpen
  \bibfield  {author} {\bibinfo {author} {\bibfnamefont {A.}~\bibnamefont
  {Krasnok}}, \bibinfo {author} {\bibfnamefont {D.}~\bibnamefont {Baranov}},
  \bibinfo {author} {\bibfnamefont {H.}~\bibnamefont {Li}}, \bibinfo {author}
  {\bibfnamefont {M.-A.}\ \bibnamefont {Miri}}, \bibinfo {author}
  {\bibfnamefont {F.}~\bibnamefont {Monticone}},\ and\ \bibinfo {author}
  {\bibfnamefont {A.}~\bibnamefont {Al{\`u}}},\ }\bibfield  {title} {\bibinfo
  {title} {Anomalies in light scattering},\ }\href
  {https://doi.org/10.1364/AOP.11.000892} {\bibfield  {journal} {\bibinfo
  {journal} {Advances in Optics and Photonics}\ }\textbf {\bibinfo {volume}
  {11}},\ \bibinfo {pages} {892} (\bibinfo {year} {2019})}\BibitemShut
  {NoStop}%
\bibitem [{\citenamefont {Chong}\ \emph {et~al.}(2010)\citenamefont {Chong},
  \citenamefont {Ge}, \citenamefont {Cao},\ and\ \citenamefont
  {Stone}}]{Chong2010}%
  \BibitemOpen
  \bibfield  {author} {\bibinfo {author} {\bibfnamefont {Y.~D.}\ \bibnamefont
  {Chong}}, \bibinfo {author} {\bibfnamefont {L.}~\bibnamefont {Ge}}, \bibinfo
  {author} {\bibfnamefont {H.}~\bibnamefont {Cao}},\ and\ \bibinfo {author}
  {\bibfnamefont {A.~D.}\ \bibnamefont {Stone}},\ }\bibfield  {title} {\bibinfo
  {title} {Coherent perfect absorbers: Time-reversed lasers},\ }\href
  {https://doi.org/10.1103/PhysRevLett.105.053901} {\bibfield  {journal}
  {\bibinfo  {journal} {Physical Review Letters}\ }\textbf {\bibinfo {volume}
  {105}},\ \bibinfo {eid} {053901} (\bibinfo {year} {2010})}\BibitemShut
  {NoStop}%
\bibitem [{\citenamefont {Baranov}\ \emph
  {et~al.}(2017{\natexlab{a}})\citenamefont {Baranov}, \citenamefont {Krasnok},
  \citenamefont {Shegai}, \citenamefont {Al{\`u}},\ and\ \citenamefont
  {Chong}}]{BaranovNRM2017}%
  \BibitemOpen
  \bibfield  {author} {\bibinfo {author} {\bibfnamefont {D.~G.}\ \bibnamefont
  {Baranov}}, \bibinfo {author} {\bibfnamefont {A.}~\bibnamefont {Krasnok}},
  \bibinfo {author} {\bibfnamefont {T.}~\bibnamefont {Shegai}}, \bibinfo
  {author} {\bibfnamefont {A.}~\bibnamefont {Al{\`u}}},\ and\ \bibinfo {author}
  {\bibfnamefont {Y.}~\bibnamefont {Chong}},\ }\bibfield  {title} {\bibinfo
  {title} {Coherent perfect absorbers: linear control of light with light},\
  }\href {https://doi.org/10.1038/natrevmats.2017.64} {\bibfield  {journal}
  {\bibinfo  {journal} {Nature Reviews Materials}\ }\textbf {\bibinfo {volume}
  {2}},\ \bibinfo {eid} {17064} (\bibinfo {year}
  {2017}{\natexlab{a}})}\BibitemShut {NoStop}%
\bibitem [{\citenamefont {Sweeney}\ \emph {et~al.}(2020)\citenamefont
  {Sweeney}, \citenamefont {Hsu},\ and\ \citenamefont {Stone}}]{Sweeney2020}%
  \BibitemOpen
  \bibfield  {author} {\bibinfo {author} {\bibfnamefont {W.~R.}\ \bibnamefont
  {Sweeney}}, \bibinfo {author} {\bibfnamefont {C.~W.}\ \bibnamefont {Hsu}},\
  and\ \bibinfo {author} {\bibfnamefont {A.~D.}\ \bibnamefont {Stone}},\
  }\bibfield  {title} {\bibinfo {title} {Theory of reflectionless scattering
  modes},\ }\href {https://doi.org/10.1103/PhysRevA.102.063511} {\bibfield
  {journal} {\bibinfo  {journal} {Physical Review A}\ }\textbf {\bibinfo
  {volume} {102}},\ \bibinfo {eid} {063511} (\bibinfo {year}
  {2020})}\BibitemShut {NoStop}%
\bibitem [{\citenamefont {Stone}\ \emph {et~al.}(2021)\citenamefont {Stone},
  \citenamefont {Sweeney}, \citenamefont {Hsu}, \citenamefont {Wisal},\ and\
  \citenamefont {Wang}}]{Stone2021}%
  \BibitemOpen
  \bibfield  {author} {\bibinfo {author} {\bibfnamefont {A.~D.}\ \bibnamefont
  {Stone}}, \bibinfo {author} {\bibfnamefont {W.~R.}\ \bibnamefont {Sweeney}},
  \bibinfo {author} {\bibfnamefont {C.~W.}\ \bibnamefont {Hsu}}, \bibinfo
  {author} {\bibfnamefont {K.}~\bibnamefont {Wisal}},\ and\ \bibinfo {author}
  {\bibfnamefont {Z.}~\bibnamefont {Wang}},\ }\bibfield  {title} {\bibinfo
  {title} {Reflectionless excitation of arbitrary photonic structures: A
  general theory},\ }\href {https://doi.org/10.1515/nanoph-2020-0403}
  {\bibfield  {journal} {\bibinfo  {journal} {Nanophotonics}\ }\textbf
  {\bibinfo {volume} {10}},\ \bibinfo {pages} {343} (\bibinfo {year}
  {2021})}\BibitemShut {NoStop}%
\bibitem [{\citenamefont {Sweeney}\ \emph {et~al.}(2019)\citenamefont
  {Sweeney}, \citenamefont {Hsu}, \citenamefont {Rotter},\ and\ \citenamefont
  {Stone}}]{Sweeney2019}%
  \BibitemOpen
  \bibfield  {author} {\bibinfo {author} {\bibfnamefont {W.~R.}\ \bibnamefont
  {Sweeney}}, \bibinfo {author} {\bibfnamefont {C.~W.}\ \bibnamefont {Hsu}},
  \bibinfo {author} {\bibfnamefont {S.}~\bibnamefont {Rotter}},\ and\ \bibinfo
  {author} {\bibfnamefont {A.~D.}\ \bibnamefont {Stone}},\ }\bibfield  {title}
  {\bibinfo {title} {Perfectly absorbing exceptional points and chiral
  absorbers},\ }\href {https://doi.org/10.1103/PhysRevLett.122.093901}
  {\bibfield  {journal} {\bibinfo  {journal} {Physical Review Letters}\
  }\textbf {\bibinfo {volume} {122}},\ \bibinfo {eid} {093901} (\bibinfo {year}
  {2019})}\BibitemShut {NoStop}%
\bibitem [{\citenamefont {Wang}\ \emph {et~al.}(2021)\citenamefont {Wang},
  \citenamefont {Sweeney}, \citenamefont {Stone},\ and\ \citenamefont
  {Yang}}]{Wang2021}%
  \BibitemOpen
  \bibfield  {author} {\bibinfo {author} {\bibfnamefont {C.}~\bibnamefont
  {Wang}}, \bibinfo {author} {\bibfnamefont {W.~R.}\ \bibnamefont {Sweeney}},
  \bibinfo {author} {\bibfnamefont {A.~D.}\ \bibnamefont {Stone}},\ and\
  \bibinfo {author} {\bibfnamefont {L.}~\bibnamefont {Yang}},\ }\bibfield
  {title} {\bibinfo {title} {Coherent perfect absorption at an exceptional
  point},\ }\href {https://doi.org/10.1126/science.abj1028} {\bibfield
  {journal} {\bibinfo  {journal} {Science}\ }\textbf {\bibinfo {volume}
  {373}},\ \bibinfo {pages} {1261} (\bibinfo {year} {2021})}\BibitemShut
  {NoStop}%
\bibitem [{\citenamefont {Ferise}\ \emph {et~al.}(2022)\citenamefont {Ferise},
  \citenamefont {del Hougne}, \citenamefont {F{\'e}lix}, \citenamefont
  {Pagneux},\ and\ \citenamefont {Davy}}]{Ferise2022}%
  \BibitemOpen
  \bibfield  {author} {\bibinfo {author} {\bibfnamefont {C.}~\bibnamefont
  {Ferise}}, \bibinfo {author} {\bibfnamefont {P.}~\bibnamefont {del Hougne}},
  \bibinfo {author} {\bibfnamefont {S.}~\bibnamefont {F{\'e}lix}}, \bibinfo
  {author} {\bibfnamefont {V.}~\bibnamefont {Pagneux}},\ and\ \bibinfo {author}
  {\bibfnamefont {M.}~\bibnamefont {Davy}},\ }\bibfield  {title} {\bibinfo
  {title} {Exceptional points of {PT}-symmetric reflectionless states in
  complex scattering systems},\ }\href
  {https://doi.org/10.1103/PhysRevLett.128.203904} {\bibfield  {journal}
  {\bibinfo  {journal} {Physical Review Letters}\ }\textbf {\bibinfo {volume}
  {128}},\ \bibinfo {eid} {203904} (\bibinfo {year} {2022})}\BibitemShut
  {NoStop}%
\bibitem [{\citenamefont {Kravets}\ \emph {et~al.}(2013)\citenamefont
  {Kravets}, \citenamefont {Schedin}, \citenamefont {Jalil}, \citenamefont
  {Britnell}, \citenamefont {Gorbachev}, \citenamefont {Ansell}, \citenamefont
  {Thackray}, \citenamefont {Novoselov}, \citenamefont {Geim}, \citenamefont
  {Kabashin},\ and\ \citenamefont {Grigorenko}}]{Kravets2013}%
  \BibitemOpen
  \bibfield  {author} {\bibinfo {author} {\bibfnamefont {V.~G.}\ \bibnamefont
  {Kravets}}, \bibinfo {author} {\bibfnamefont {F.}~\bibnamefont {Schedin}},
  \bibinfo {author} {\bibfnamefont {R.}~\bibnamefont {Jalil}}, \bibinfo
  {author} {\bibfnamefont {L.}~\bibnamefont {Britnell}}, \bibinfo {author}
  {\bibfnamefont {R.~V.}\ \bibnamefont {Gorbachev}}, \bibinfo {author}
  {\bibfnamefont {D.}~\bibnamefont {Ansell}}, \bibinfo {author} {\bibfnamefont
  {B.}~\bibnamefont {Thackray}}, \bibinfo {author} {\bibfnamefont {K.~S.}\
  \bibnamefont {Novoselov}}, \bibinfo {author} {\bibfnamefont {A.~K.}\
  \bibnamefont {Geim}}, \bibinfo {author} {\bibfnamefont {A.~V.}\ \bibnamefont
  {Kabashin}},\ and\ \bibinfo {author} {\bibfnamefont {A.~N.}\ \bibnamefont
  {Grigorenko}},\ }\bibfield  {title} {\bibinfo {title} {Singular phase
  nano-optics in plasmonic metamaterials for label-free single-molecule
  detection},\ }\href {https://doi.org/10.1038/nmat3537} {\bibfield  {journal}
  {\bibinfo  {journal} {Nature Materials}\ }\textbf {\bibinfo {volume} {12}},\
  \bibinfo {pages} {304} (\bibinfo {year} {2013})}\BibitemShut {NoStop}%
\bibitem [{\citenamefont {Soori}\ \emph {et~al.}(2026)\citenamefont {Soori},
  \citenamefont {Khanna},\ and\ \citenamefont {Sen}}]{Soori2026}%
  \BibitemOpen
  \bibfield  {author} {\bibinfo {author} {\bibfnamefont {A.}~\bibnamefont
  {Soori}}, \bibinfo {author} {\bibfnamefont {U.}~\bibnamefont {Khanna}},\ and\
  \bibinfo {author} {\bibfnamefont {D.}~\bibnamefont {Sen}},\ }\href@noop {}
  {\bibinfo {title} {Topological zero-reflection points in multi-terminal
  quantum wire junctions}} (\bibinfo {year} {2026}),\ \Eprint
  {https://arxiv.org/abs/2606.31586} {arXiv:2606.31586} \BibitemShut {NoStop}%
\bibitem [{\citenamefont {Franca}\ \emph {et~al.}(2022)\citenamefont {Franca},
  \citenamefont {K{\"o}nye}, \citenamefont {Hassler}, \citenamefont {van~den
  Brink},\ and\ \citenamefont {Fulga}}]{Franca2022}%
  \BibitemOpen
  \bibfield  {author} {\bibinfo {author} {\bibfnamefont {S.}~\bibnamefont
  {Franca}}, \bibinfo {author} {\bibfnamefont {V.}~\bibnamefont {K{\"o}nye}},
  \bibinfo {author} {\bibfnamefont {F.}~\bibnamefont {Hassler}}, \bibinfo
  {author} {\bibfnamefont {J.}~\bibnamefont {van~den Brink}},\ and\ \bibinfo
  {author} {\bibfnamefont {C.}~\bibnamefont {Fulga}},\ }\bibfield  {title}
  {\bibinfo {title} {Non-{H}ermitian physics without gain or loss: The skin
  effect of reflected waves},\ }\href
  {https://doi.org/10.1103/PhysRevLett.129.086601} {\bibfield  {journal}
  {\bibinfo  {journal} {Physical Review Letters}\ }\textbf {\bibinfo {volume}
  {129}},\ \bibinfo {eid} {086601} (\bibinfo {year} {2022})}\BibitemShut
  {NoStop}%
\bibitem [{\citenamefont {Soley}\ \emph {et~al.}(2023)\citenamefont {Soley},
  \citenamefont {Bender},\ and\ \citenamefont {Stone}}]{Soley2023}%
  \BibitemOpen
  \bibfield  {author} {\bibinfo {author} {\bibfnamefont {M.~B.}\ \bibnamefont
  {Soley}}, \bibinfo {author} {\bibfnamefont {C.~M.}\ \bibnamefont {Bender}},\
  and\ \bibinfo {author} {\bibfnamefont {A.~D.}\ \bibnamefont {Stone}},\
  }\bibfield  {title} {\bibinfo {title} {Experimentally realizable
  {$\mathcal{PT}$} phase transitions in reflectionless quantum scattering},\
  }\href {https://doi.org/10.1103/PhysRevLett.130.250404} {\bibfield  {journal}
  {\bibinfo  {journal} {Physical Review Letters}\ }\textbf {\bibinfo {volume}
  {130}},\ \bibinfo {eid} {250404} (\bibinfo {year} {2023})}\BibitemShut
  {NoStop}%
\bibitem [{\citenamefont {Ohnmacht}\ \emph {et~al.}(2025)\citenamefont
  {Ohnmacht}, \citenamefont {Wilhelm},\ and\ \citenamefont
  {Belzig}}]{Ohnmacht2025}%
  \BibitemOpen
  \bibfield  {author} {\bibinfo {author} {\bibfnamefont {D.~C.}\ \bibnamefont
  {Ohnmacht}}, \bibinfo {author} {\bibfnamefont {V.}~\bibnamefont {Wilhelm}},\
  and\ \bibinfo {author} {\bibfnamefont {W.}~\bibnamefont {Belzig}},\
  }\href@noop {} {\bibinfo {title} {Reflectionless modes as a source of {Weyl}
  nodes in multiterminal {Josephson} junctions}} (\bibinfo {year} {2025}),\
  \Eprint {https://arxiv.org/abs/2503.10874} {arXiv:2503.10874} \BibitemShut
  {NoStop}%
\bibitem [{\citenamefont {Mahaux}\ and\ \citenamefont
  {Weidenm{\"u}ller}(1969)}]{Mahaux1969}%
  \BibitemOpen
  \bibfield  {author} {\bibinfo {author} {\bibfnamefont {C.}~\bibnamefont
  {Mahaux}}\ and\ \bibinfo {author} {\bibfnamefont {H.~A.}\ \bibnamefont
  {Weidenm{\"u}ller}},\ }\href@noop {} {\emph {\bibinfo {title} {Shell-Model
  Approach to Nuclear Reactions}}}\ (\bibinfo  {publisher} {North-Holland},\
  \bibinfo {address} {Amsterdam},\ \bibinfo {year} {1969})\BibitemShut
  {NoStop}%
\bibitem [{SM()}]{SM}%
  \BibitemOpen
  \href@noop {} {}\bibinfo {note} {See Supplemental Material for the
  derivations, the tests of Eq.~(2), the extensions to other symmetry classes
  and to loss, the trivial zero-bias peak, the Kitaev chain, and the models and
  numerical methods}\BibitemShut {NoStop}%
\bibitem [{\citenamefont {Li}\ \emph {et~al.}(2017)\citenamefont {Li},
  \citenamefont {Suwunnarat}, \citenamefont {Fleischmann}, \citenamefont
  {Schanz},\ and\ \citenamefont {Kottos}}]{Li2017}%
  \BibitemOpen
  \bibfield  {author} {\bibinfo {author} {\bibfnamefont {H.}~\bibnamefont
  {Li}}, \bibinfo {author} {\bibfnamefont {S.}~\bibnamefont {Suwunnarat}},
  \bibinfo {author} {\bibfnamefont {R.}~\bibnamefont {Fleischmann}}, \bibinfo
  {author} {\bibfnamefont {H.}~\bibnamefont {Schanz}},\ and\ \bibinfo {author}
  {\bibfnamefont {T.}~\bibnamefont {Kottos}},\ }\bibfield  {title} {\bibinfo
  {title} {Random matrix theory approach to chaotic coherent perfect
  absorbers},\ }\href {https://doi.org/10.1103/PhysRevLett.118.044101}
  {\bibfield  {journal} {\bibinfo  {journal} {Physical Review Letters}\
  }\textbf {\bibinfo {volume} {118}},\ \bibinfo {eid} {044101} (\bibinfo {year}
  {2017})}\BibitemShut {NoStop}%
\bibitem [{\citenamefont {Fyodorov}\ \emph {et~al.}(2017)\citenamefont
  {Fyodorov}, \citenamefont {Suwunnarat},\ and\ \citenamefont
  {Kottos}}]{Fyodorov2017}%
  \BibitemOpen
  \bibfield  {author} {\bibinfo {author} {\bibfnamefont {Y.~V.}\ \bibnamefont
  {Fyodorov}}, \bibinfo {author} {\bibfnamefont {S.}~\bibnamefont
  {Suwunnarat}},\ and\ \bibinfo {author} {\bibfnamefont {T.}~\bibnamefont
  {Kottos}},\ }\bibfield  {title} {\bibinfo {title} {Distribution of zeros of
  the {S}-matrix of chaotic cavities with localized losses and coherent perfect
  absorption: Non-perturbative results},\ }\href
  {https://doi.org/10.1088/1751-8121/aa793a} {\bibfield  {journal} {\bibinfo
  {journal} {Journal of Physics A: Mathematical and Theoretical}\ }\textbf
  {\bibinfo {volume} {50}},\ \bibinfo {eid} {30LT01} (\bibinfo {year}
  {2017})}\BibitemShut {NoStop}%
\bibitem [{\citenamefont {Nilsson}\ \emph {et~al.}(2008)\citenamefont
  {Nilsson}, \citenamefont {Akhmerov},\ and\ \citenamefont
  {Beenakker}}]{Nilsson2008}%
  \BibitemOpen
  \bibfield  {author} {\bibinfo {author} {\bibfnamefont {J.}~\bibnamefont
  {Nilsson}}, \bibinfo {author} {\bibfnamefont {A.~R.}\ \bibnamefont
  {Akhmerov}},\ and\ \bibinfo {author} {\bibfnamefont {C.~W.~J.}\ \bibnamefont
  {Beenakker}},\ }\bibfield  {title} {\bibinfo {title} {Splitting of a {C}ooper
  pair by a pair of {M}ajorana bound states},\ }\href
  {https://doi.org/10.1103/PhysRevLett.101.120403} {\bibfield  {journal}
  {\bibinfo  {journal} {Physical Review Letters}\ }\textbf {\bibinfo {volume}
  {101}},\ \bibinfo {eid} {120403} (\bibinfo {year} {2008})}\BibitemShut
  {NoStop}%
\bibitem [{\citenamefont {Budich}\ \emph {et~al.}(2019)\citenamefont {Budich},
  \citenamefont {Carlstr{\"o}m}, \citenamefont {Kunst},\ and\ \citenamefont
  {Bergholtz}}]{Budich2019}%
  \BibitemOpen
  \bibfield  {author} {\bibinfo {author} {\bibfnamefont {J.~C.}\ \bibnamefont
  {Budich}}, \bibinfo {author} {\bibfnamefont {J.}~\bibnamefont
  {Carlstr{\"o}m}}, \bibinfo {author} {\bibfnamefont {F.~K.}\ \bibnamefont
  {Kunst}},\ and\ \bibinfo {author} {\bibfnamefont {E.~J.}\ \bibnamefont
  {Bergholtz}},\ }\bibfield  {title} {\bibinfo {title} {Symmetry-protected
  nodal phases in non-{Hermitian} systems},\ }\href
  {https://doi.org/10.1103/PhysRevB.99.041406} {\bibfield  {journal} {\bibinfo
  {journal} {Physical Review B}\ }\textbf {\bibinfo {volume} {99}},\ \bibinfo
  {eid} {041406} (\bibinfo {year} {2019})}\BibitemShut {NoStop}%
\bibitem [{\citenamefont {Okugawa}\ and\ \citenamefont
  {Yokoyama}(2019)}]{Okugawa2019}%
  \BibitemOpen
  \bibfield  {author} {\bibinfo {author} {\bibfnamefont {R.}~\bibnamefont
  {Okugawa}}\ and\ \bibinfo {author} {\bibfnamefont {T.}~\bibnamefont
  {Yokoyama}},\ }\bibfield  {title} {\bibinfo {title} {Topological exceptional
  surfaces in non-{Hermitian} systems with parity-time and parity-particle-hole
  symmetries},\ }\href {https://doi.org/10.1103/PhysRevB.99.041202} {\bibfield
  {journal} {\bibinfo  {journal} {Physical Review B}\ }\textbf {\bibinfo
  {volume} {99}},\ \bibinfo {eid} {041202} (\bibinfo {year}
  {2019})}\BibitemShut {NoStop}%
\bibitem [{\citenamefont {Kawabata}\ \emph {et~al.}(2019)\citenamefont
  {Kawabata}, \citenamefont {Shiozaki}, \citenamefont {Ueda},\ and\
  \citenamefont {Sato}}]{Kawabata2019}%
  \BibitemOpen
  \bibfield  {author} {\bibinfo {author} {\bibfnamefont {K.}~\bibnamefont
  {Kawabata}}, \bibinfo {author} {\bibfnamefont {K.}~\bibnamefont {Shiozaki}},
  \bibinfo {author} {\bibfnamefont {M.}~\bibnamefont {Ueda}},\ and\ \bibinfo
  {author} {\bibfnamefont {M.}~\bibnamefont {Sato}},\ }\bibfield  {title}
  {\bibinfo {title} {Symmetry and topology in non-{Hermitian} physics},\ }\href
  {https://doi.org/10.1103/PhysRevX.9.041015} {\bibfield  {journal} {\bibinfo
  {journal} {Phys. Rev. X}\ }\textbf {\bibinfo {volume} {9}},\ \bibinfo {pages}
  {041015} (\bibinfo {year} {2019})}\BibitemShut {NoStop}%
\bibitem [{\citenamefont {Butterworth}(1930)}]{Butterworth1930}%
  \BibitemOpen
  \bibfield  {author} {\bibinfo {author} {\bibfnamefont {S.}~\bibnamefont
  {Butterworth}},\ }\bibfield  {title} {\bibinfo {title} {On the theory of
  filter amplifiers},\ }\href@noop {} {\bibfield  {journal} {\bibinfo
  {journal} {Experimental Wireless and the Wireless Engineer}\ }\textbf
  {\bibinfo {volume} {7}},\ \bibinfo {pages} {536} (\bibinfo {year}
  {1930})}\BibitemShut {NoStop}%
\bibitem [{\citenamefont {Rosdahl}\ \emph {et~al.}(2018)\citenamefont
  {Rosdahl}, \citenamefont {Vuik}, \citenamefont {Kjaergaard},\ and\
  \citenamefont {Akhmerov}}]{Rosdahl2018}%
  \BibitemOpen
  \bibfield  {author} {\bibinfo {author} {\bibfnamefont {T.~{\"O}.}\
  \bibnamefont {Rosdahl}}, \bibinfo {author} {\bibfnamefont {A.}~\bibnamefont
  {Vuik}}, \bibinfo {author} {\bibfnamefont {M.}~\bibnamefont {Kjaergaard}},\
  and\ \bibinfo {author} {\bibfnamefont {A.~R.}\ \bibnamefont {Akhmerov}},\
  }\bibfield  {title} {\bibinfo {title} {{A}ndreev rectifier: A nonlocal
  conductance signature of topological phase transitions},\ }\href
  {https://doi.org/10.1103/PhysRevB.97.045421} {\bibfield  {journal} {\bibinfo
  {journal} {Physical Review B}\ }\textbf {\bibinfo {volume} {97}},\ \bibinfo
  {eid} {045421} (\bibinfo {year} {2018})}\BibitemShut {NoStop}%
\bibitem [{\citenamefont {Danon}\ \emph {et~al.}(2020)\citenamefont {Danon},
  \citenamefont {Hellenes}, \citenamefont {Hansen}, \citenamefont {Casparis},
  \citenamefont {Higginbotham},\ and\ \citenamefont {Flensberg}}]{Danon2020}%
  \BibitemOpen
  \bibfield  {author} {\bibinfo {author} {\bibfnamefont {J.}~\bibnamefont
  {Danon}}, \bibinfo {author} {\bibfnamefont {A.~B.}\ \bibnamefont {Hellenes}},
  \bibinfo {author} {\bibfnamefont {E.~B.}\ \bibnamefont {Hansen}}, \bibinfo
  {author} {\bibfnamefont {L.}~\bibnamefont {Casparis}}, \bibinfo {author}
  {\bibfnamefont {A.~P.}\ \bibnamefont {Higginbotham}},\ and\ \bibinfo {author}
  {\bibfnamefont {K.}~\bibnamefont {Flensberg}},\ }\bibfield  {title} {\bibinfo
  {title} {Nonlocal conductance spectroscopy of {A}ndreev bound states:
  Symmetry relations and {BCS} charges},\ }\href
  {https://doi.org/10.1103/PhysRevLett.124.036801} {\bibfield  {journal}
  {\bibinfo  {journal} {Physical Review Letters}\ }\textbf {\bibinfo {volume}
  {124}},\ \bibinfo {eid} {036801} (\bibinfo {year} {2020})}\BibitemShut
  {NoStop}%
\bibitem [{\citenamefont {Wimmer}\ \emph {et~al.}(2011)\citenamefont {Wimmer},
  \citenamefont {Akhmerov}, \citenamefont {Dahlhaus},\ and\ \citenamefont
  {Beenakker}}]{Wimmer2011}%
  \BibitemOpen
  \bibfield  {author} {\bibinfo {author} {\bibfnamefont {M.}~\bibnamefont
  {Wimmer}}, \bibinfo {author} {\bibfnamefont {A.~R.}\ \bibnamefont
  {Akhmerov}}, \bibinfo {author} {\bibfnamefont {J.~P.}\ \bibnamefont
  {Dahlhaus}},\ and\ \bibinfo {author} {\bibfnamefont {C.~W.~J.}\ \bibnamefont
  {Beenakker}},\ }\bibfield  {title} {\bibinfo {title} {Quantum point contact
  as a probe of a topological superconductor},\ }\href
  {https://doi.org/10.1088/1367-2630/13/5/053016} {\bibfield  {journal}
  {\bibinfo  {journal} {New Journal of Physics}\ }\textbf {\bibinfo {volume}
  {13}},\ \bibinfo {eid} {053016} (\bibinfo {year} {2011})}\BibitemShut
  {NoStop}%
\bibitem [{\citenamefont {Das~Sarma}\ \emph {et~al.}(2016)\citenamefont
  {Das~Sarma}, \citenamefont {Nag},\ and\ \citenamefont {Sau}}]{DasSarma2016}%
  \BibitemOpen
  \bibfield  {author} {\bibinfo {author} {\bibfnamefont {S.}~\bibnamefont
  {Das~Sarma}}, \bibinfo {author} {\bibfnamefont {A.}~\bibnamefont {Nag}},\
  and\ \bibinfo {author} {\bibfnamefont {J.~D.}\ \bibnamefont {Sau}},\
  }\bibfield  {title} {\bibinfo {title} {How to infer non-{Abelian} statistics
  and topological visibility from tunneling conductance properties of realistic
  {Majorana} nanowires},\ }\href {https://doi.org/10.1103/PhysRevB.94.035143}
  {\bibfield  {journal} {\bibinfo  {journal} {Physical Review B}\ }\textbf
  {\bibinfo {volume} {94}},\ \bibinfo {eid} {035143} (\bibinfo {year}
  {2016})}\BibitemShut {NoStop}%
\bibitem [{\citenamefont {Groth}\ \emph {et~al.}(2014)\citenamefont {Groth},
  \citenamefont {Wimmer}, \citenamefont {Akhmerov},\ and\ \citenamefont
  {Waintal}}]{Groth2014}%
  \BibitemOpen
  \bibfield  {author} {\bibinfo {author} {\bibfnamefont {C.~W.}\ \bibnamefont
  {Groth}}, \bibinfo {author} {\bibfnamefont {M.}~\bibnamefont {Wimmer}},
  \bibinfo {author} {\bibfnamefont {A.~R.}\ \bibnamefont {Akhmerov}},\ and\
  \bibinfo {author} {\bibfnamefont {X.}~\bibnamefont {Waintal}},\ }\bibfield
  {title} {\bibinfo {title} {Kwant: a software package for quantum transport},\
  }\href {https://doi.org/10.1088/1367-2630/16/6/063065} {\bibfield  {journal}
  {\bibinfo  {journal} {New Journal of Physics}\ }\textbf {\bibinfo {volume}
  {16}},\ \bibinfo {eid} {063065} (\bibinfo {year} {2014})}\BibitemShut
  {NoStop}%
\bibitem [{\citenamefont {Das~Sarma}\ \emph {et~al.}(2012)\citenamefont
  {Das~Sarma}, \citenamefont {Sau},\ and\ \citenamefont
  {Stanescu}}]{DasSarma2012}%
  \BibitemOpen
  \bibfield  {author} {\bibinfo {author} {\bibfnamefont {S.}~\bibnamefont
  {Das~Sarma}}, \bibinfo {author} {\bibfnamefont {J.~D.}\ \bibnamefont {Sau}},\
  and\ \bibinfo {author} {\bibfnamefont {T.~D.}\ \bibnamefont {Stanescu}},\
  }\bibfield  {title} {\bibinfo {title} {Splitting of the zero-bias conductance
  peak as smoking gun evidence for the existence of the {Majorana} mode in a
  superconductor-semiconductor nanowire},\ }\href
  {https://doi.org/10.1103/PhysRevB.86.220506} {\bibfield  {journal} {\bibinfo
  {journal} {Physical Review B}\ }\textbf {\bibinfo {volume} {86}},\ \bibinfo
  {eid} {220506} (\bibinfo {year} {2012})}\BibitemShut {NoStop}%
\bibitem [{\citenamefont {Albrecht}\ \emph {et~al.}(2016)\citenamefont
  {Albrecht}, \citenamefont {Higginbotham}, \citenamefont {Madsen},
  \citenamefont {Kuemmeth}, \citenamefont {Jespersen}, \citenamefont
  {Nyg{\aa}rd}, \citenamefont {Krogstrup},\ and\ \citenamefont
  {Marcus}}]{Albrecht2016}%
  \BibitemOpen
  \bibfield  {author} {\bibinfo {author} {\bibfnamefont {S.~M.}\ \bibnamefont
  {Albrecht}}, \bibinfo {author} {\bibfnamefont {A.~P.}\ \bibnamefont
  {Higginbotham}}, \bibinfo {author} {\bibfnamefont {M.}~\bibnamefont
  {Madsen}}, \bibinfo {author} {\bibfnamefont {F.}~\bibnamefont {Kuemmeth}},
  \bibinfo {author} {\bibfnamefont {T.~S.}\ \bibnamefont {Jespersen}}, \bibinfo
  {author} {\bibfnamefont {J.}~\bibnamefont {Nyg{\aa}rd}}, \bibinfo {author}
  {\bibfnamefont {P.}~\bibnamefont {Krogstrup}},\ and\ \bibinfo {author}
  {\bibfnamefont {C.~M.}\ \bibnamefont {Marcus}},\ }\bibfield  {title}
  {\bibinfo {title} {Exponential protection of zero modes in {Majorana}
  islands},\ }\href {https://doi.org/10.1038/nature17162} {\bibfield  {journal}
  {\bibinfo  {journal} {Nature}\ }\textbf {\bibinfo {volume} {531}},\ \bibinfo
  {pages} {206} (\bibinfo {year} {2016})}\BibitemShut {NoStop}%
\bibitem [{\citenamefont {Chen}\ \emph
  {et~al.}(2021{\natexlab{a}})\citenamefont {Chen}, \citenamefont {Anlage},\
  and\ \citenamefont {Fyodorov}}]{Chen2021PRL}%
  \BibitemOpen
  \bibfield  {author} {\bibinfo {author} {\bibfnamefont {L.}~\bibnamefont
  {Chen}}, \bibinfo {author} {\bibfnamefont {S.~M.}\ \bibnamefont {Anlage}},\
  and\ \bibinfo {author} {\bibfnamefont {Y.~V.}\ \bibnamefont {Fyodorov}},\
  }\bibfield  {title} {\bibinfo {title} {Statistics of complex {W}igner time
  delays as a counter of {S}-matrix poles: Theory and experiment},\ }\href
  {https://doi.org/10.1103/PhysRevLett.127.204101} {\bibfield  {journal}
  {\bibinfo  {journal} {Physical Review Letters}\ }\textbf {\bibinfo {volume}
  {127}},\ \bibinfo {eid} {204101} (\bibinfo {year}
  {2021}{\natexlab{a}})}\BibitemShut {NoStop}%
\bibitem [{\citenamefont {Chen}\ \emph
  {et~al.}(2021{\natexlab{b}})\citenamefont {Chen}, \citenamefont {Anlage},\
  and\ \citenamefont {Fyodorov}}]{Chen2021PRE}%
  \BibitemOpen
  \bibfield  {author} {\bibinfo {author} {\bibfnamefont {L.}~\bibnamefont
  {Chen}}, \bibinfo {author} {\bibfnamefont {S.~M.}\ \bibnamefont {Anlage}},\
  and\ \bibinfo {author} {\bibfnamefont {Y.~V.}\ \bibnamefont {Fyodorov}},\
  }\bibfield  {title} {\bibinfo {title} {Generalization of {W}igner time delay
  to subunitary scattering systems},\ }\href
  {https://doi.org/10.1103/PhysRevE.103.L050203} {\bibfield  {journal}
  {\bibinfo  {journal} {Physical Review E}\ }\textbf {\bibinfo {volume}
  {103}},\ \bibinfo {eid} {L050203} (\bibinfo {year}
  {2021}{\natexlab{b}})}\BibitemShut {NoStop}%
\bibitem [{\citenamefont {Baranov}\ \emph
  {et~al.}(2017{\natexlab{b}})\citenamefont {Baranov}, \citenamefont
  {Krasnok},\ and\ \citenamefont {Al{\`u}}}]{Baranov2017}%
  \BibitemOpen
  \bibfield  {author} {\bibinfo {author} {\bibfnamefont {D.~G.}\ \bibnamefont
  {Baranov}}, \bibinfo {author} {\bibfnamefont {A.}~\bibnamefont {Krasnok}},\
  and\ \bibinfo {author} {\bibfnamefont {A.}~\bibnamefont {Al{\`u}}},\
  }\bibfield  {title} {\bibinfo {title} {Coherent virtual absorption based on
  complex zero excitation for ideal light capturing},\ }\href
  {https://doi.org/10.1364/OPTICA.4.001457} {\bibfield  {journal} {\bibinfo
  {journal} {Optica}\ }\textbf {\bibinfo {volume} {4}},\ \bibinfo {pages}
  {1457} (\bibinfo {year} {2017}{\natexlab{b}})}\BibitemShut {NoStop}%
\bibitem [{\citenamefont {Kim}\ \emph {et~al.}(2025)\citenamefont {Kim},
  \citenamefont {Krasnok},\ and\ \citenamefont {Al{\`u}}}]{KimReview2025}%
  \BibitemOpen
  \bibfield  {author} {\bibinfo {author} {\bibfnamefont {S.}~\bibnamefont
  {Kim}}, \bibinfo {author} {\bibfnamefont {A.}~\bibnamefont {Krasnok}},\ and\
  \bibinfo {author} {\bibfnamefont {A.}~\bibnamefont {Al{\`u}}},\ }\bibfield
  {title} {\bibinfo {title} {Complex-frequency excitations in photonics and
  wave physics},\ }\href {https://doi.org/10.1126/science.ado4128} {\bibfield
  {journal} {\bibinfo  {journal} {Science}\ }\textbf {\bibinfo {volume}
  {387}},\ \bibinfo {eid} {eado4128} (\bibinfo {year} {2025})}\BibitemShut
  {NoStop}%
\end{thebibliography}%

\end{document}